\documentclass[iop]{emulateapj} 
\usepackage{amsmath}
\usepackage{amssymb}
\usepackage{mathbbol}
\usepackage{dsfont}
\usepackage{graphicx}
\usepackage{epstopdf}
\usepackage{pst-eps}
\usepackage{diagbox}

\DeclareGraphicsExtensions{.eps,.eps.gz,.epsi}

\newcommand{\teff}{\mbox{$T_{\rm eff}$}} 
\newcommand{\logg}{{\rm{log}~$g$}}
\newcommand{\feh}{{\rm [Fe/H]}} 
\newcommand{\ebv}{$E(B-V)$}

\newcommand{\ebprp}{$E(BP-RP)$} 

\shorttitle{Photometric recalibration of Stripe 82 to a few mmag precision}

\begin{document}

\title
{Photometric recalibration of the SDSS Stripe 82 to a few milimagnitude precision with the stellar color regression method and {\it Gaia} EDR3}

\author
{Bowen Huang\altaffilmark{1},
Haibo Yuan\altaffilmark{1} }
\altaffiltext{1}{Department of Astronomy, Beijing Normal University, Beijing 100875, P. R. China; email: yuanhb@bnu.edu.cn}

\journalinfo{ApJS Accepted}
\submitted{(Received October 9, 2021; Revised December 13, 2021; Accepted December 23, 2021)}

\begin{abstract}
By combining spectroscopic data from the LAMOST DR7, SDSS DR12, and corrected photometric data from the {\it Gaia} EDR3, we apply the Stellar Color Regression (SCR; Yuan et al. 2015a) method to recalibrate the SDSS Stripe 82 standard stars catalog of Ivezi{\'c} et al. (2007). 
With a total number of about 30,000 spectroscopically targeted stars, we have mapped out the relatively large and strongly correlated photometric zero-point errors present in the catalog, $\sim$2.5 per cent in the $u$ band and $\sim$ 1 per cent in the $griz$ bands. Our study also confirms some small but significant magnitude dependence errors in the $z$ band for some charge-coupled devices. 
Various tests show that we have achieved an internal precision of about 5 mmag in the $u$ band and about 2 mmag in the $griz$ bands, which is about 5 times better than previous results. We also apply the method to the latest version of the catalog (V4.2; Thanjavur et al. 2021), and find modest systematic calibration errors up to $\sim$ 1 per cent along the R.A. direction and smaller errors along the Dec. direction. 
The results demonstrate the power of the SCR method when combining spectroscopic data and \emph{Gaia} photometry in breaking the 1 percent precision barrier of ground-based photometric surveys. Our work paves the way for the re-calibration of the whole SDSS photometric survey and has important implications for the calibration of future surveys. Future implementations  and improvements of the SCR method under different situations are also discussed. 
\end{abstract}

\keywords{catalogs — instrumentation: photometers --  ISM: dust, extinction -- methods: data analysis -- surveys -- techniques: imaging, spectroscopic}

\section{Introduction} 
Wide-field photometric surveys play a leading role in the discovery of new celestial objects, new phenomena, and new laws in modern astronomical research. The potential discovery space and amount of information that can be 
extracted from a given photometric survey is closely related to the uniformity of its photometric calibration.
Hence, a uniform and accurate photometric calibration plays a central role in the current and next-generation wide-field imaging surveys, such as the Sloan Digital Sky Survey (SDSS; York et al. 2000), the Dark Energy Survey (DES; The Dark Energy Collaboration et al. 2005),  the Pan-STARRS1 survey (PS1; Chambers et al. 2016), the SkyMapper Southern Survey (SMSS; Wolf et al. 2018), the Stellar Abundance and Galactic Evolution survey (SAGE; Zheng et al. 2018),  the Javalambre Photometric Local Universe Survey (J-PLUS; Cenarro et al. 2019), the Southern Photometric Local Universe Survey (S-PLUS; Mendes de Oliveira et al. 2019), the Javalambre Physics of the Accelerating Universe Astrophysical Survey (J-PAS; Benitez et al. 2014).
the Vera C. Rubin Observatory Legacy Survey of Space and Time (LSST; Ivezi{\'c} et al. 2019), the Chinese Space Station Telescope (CSST; Zhan 2021), the Wide Field Survey Telescope (WFST; Lou et al. 2016), and the Multi-channel Photometric Survey Telescope (Mephisto; Er et al. 2021, in preparation).

The calibration of astronomical photometric measurements are traditionally  based on sets of standard stars 
(e.g., Landolt 1992, 2009, 2013; Stetson 2000; Clem \& Landolt 2013). 
However, due to the very limited numbers of standard stars available, 
it is very challenging to achieve calibration precision better than 1 percent for ground-based observations (e.g. Stubbs \& Tonry 2006). The challenges are mainly caused by the temporal and spatial variations of the Earth atmospheric transmission, and the difficulties in correcting for instrumental effects such as flat fielding and electronics of CCDs.
In recent years, a number of approaches have been developed for the precise calibration of wide field imaging surveys. These approaches can be roughly divided into two categories: ``hardware-driven" approaches and ``software-driven" approaches. Note that all these approaches perform ``relative" calibration (i.e., 
 establishing an internally consistent photometric calibration across the whole survey region) rather than 
 ``absolute" calibration.

The ``hardware-driven" approaches are based on better understanding of wide field imaging observations, 
including the ubercalibration method (Padmanabhan et al. 2008), the
Forward Global Calibration Method (FGCM; Burke et al. 2018), and the hypercalibration method (Finkbeiner et al. 2016). The ubercalibration method is originally developed for the SDSS but has been widely used (e.g., Schlafly et al. 2012; Liu et al. 2014; Zhou et al. 2018; Gaia Collaboration et al. 2016).  
It requires a significant amount of over-lapping observations.  
The FGCM is developed for the DES and LSST. In addition to repeated observations, 
it requires data taken with auxiliary instrumentation at the observatory, and models of the instrument and atmosphere to estimate the spatial and time variations of the passbands. 
The hypercalibration method assumes that systematic errors of different surveys are independent. Therefore,
different surveys can be used to calibrate each other.

The ``software-driven" approaches are based on better understanding of stellar colors, including the Stellar Locus Regression method (SLR; High et al. 2009), the stellar locus method (SL; L{\'o}pez-Sanjuan et al. 2019), 
and the Stellar Color Regression method (SCR; Yuan et al. 2015a). 
The position of the stellar locus was firstly used by Ivezi{\'c} et al. (2004) to 
estimate the accuracy of photometric zeropoints of the SDSS.
Assuming a universal stellar color-color locus, the SLR method can correct for effects including variations in instrumental sensitivity, in the Earth atmospheric transmission, and in the Galactic interstellar and reddening,  producing real-time well calibrated colors for both stars and galaxies.
However, the precision is limited to a few per cent, due to the variations of stellar populations and the interstellar extinction, especially in the low Galactic latitude region. The method also requires a blue filter in addition to at least two other filters to break possible degeneracy. The SLR method can only perform color 
calibration. To overcome this limitation, by using an existing well-calibrated data-set as anchors,  
the SL method can perform photometric calibration with the help of the de-reddened stellar locus. 
The method can be further improved by including the metallicity effect in colours via the 
 metallicity-dependent stellar locus (Yuan et al. 2015b; L{\'o}pez-Sanjuan et al. 2021).

With the rapid development of multi-fiber spectroscopic surveys,  e.g., SDSS/SEGUE (Yanny et al. 2009), LAMOST (Deng et al. 2012; Liu et al. 2014), APOGEE (J{\"o}nsson et al. 2020), and GALAH (De Silva et al. 2015) surveys, we have entered into the era of millions of stellar spectra. On the other hand, thanks to the modern 
template-matching based as well as data-driven stellar parameter pipelines 
(e.g., Lee et al. 2008a,b; Wu et al. 2011; Xiang et al. 2015, 2017), 
stellar atmospheric parameters ({\teff}, {\logg},  and {\feh}) 
can be determined to a very high internal precision (e.g., Niu et al. 2021a). 
Therefore, stellar colors can be accurately predicted based on the large-scale spectroscopic surveys with the star-pair technique (Yuan et al. 2013).
Taking advantage of the above fact and using millions of spectroscopically observed stars as color standards,
Yuan et al. (2015a) have proposed the spectroscopy-based SCR method to perform precise color calibrations. 
Compared to the SLR and SL methods, the SCR method fully accounts the effects of metallicity, surface gravity, 
and dust reddening on stellar colors. Applying the method to 
the SDSS Stripe 82 standard stars catalog, we have achieved a precision of 
about 5 mmag in $u - g$, 3 mmag in $g-r$, and 2 mmag in $r-i$ and $i-z$,
an improvement by a factor of two to three. The method has also been applied to the {\it Gaia} DR2 and EDR3 
(Niu et al. 2021a, b)
to correct for magnitude/color-dependent systematic errors in the {\it Gaia} colors, with a precision of about 1 millimagnitude.

With the data releases of the {\it Gaia} DR2 and EDR3 (Gaia Collaboration et al. 2016, 2018, 2021), 
accurate and homogeneous photometric data of the whole sky and with an exquisite quality are now accessible, 
reaching down to the unprecedented millimagnitude level for the G, BP, and RP passbands. With the help from
the {\it Gaia} photometry, the SCR method can accurately predict the magnitudes of stars in various bands, providing a great opportunity to break the 1 percent precision barrier of ground-based photometric surveys. Such approach 
has been applied to the second data release from the SkyMapper Southern Survey (SMSS DR2). Large zero-point offsets are detected, particularly for the gravity- and metallicity-sensitive $uv$ bands (Huang et al. 2021a).

In this work, by combining spectroscopic data from the LAMOST DR7, SDSS DR12 and corrected photometric data from the {\it Gaia} EDR3, we apply the SCR method to recalibrate the SDSS Stripe 82 standard stars catalog. 
The paper is organized as follows. In Section 2, we introduce our data.
In Section\,3, we apply the SCR method to the SDSS Stripe 82 region to recalibrate the data and check the precision of our method. 
In Section\,4, we apply the method to the latest version of the catalog (V4.2, Thanjavur et al. 2021).
The discussions and conclusions are given in Section\,5.

\section{DATASETS} 
The repeatedly scanned Stripe 82 ($|{\rm Dec.}| < 1.266\degr$, 20$^h$34$^m$ $<$ R.A. $<$ 4$^h$00$^m$) is a contiguous equatorial region of 300 deg$^2$
in the SDSS. Ivezi\'c et al (2007; I07 hereafter) delivered an 
accurate photometric catalog of about one million stars in $u,g,r,i,z$ bands, 
with an internal calibration consistency of one percent,
thus providing a practical definition of the SDSS photometric system.
The random photometric errors of the I07 catalog are below 0.01\,mag for stars brighter than 19.5, 20.5, 20.5, 20.0, and 18.5 in $u$, $g$, $r$, $i$, and $z$ bands, respectively. 

\subsection{LAMOST Data Release 7} 
For the recalibration of Stripe 82, we use spectroscopic information from LAMOST Data Release 7 (Luo et al. 2015). The Large Sky Area Multi-Object Fiber Spectroscopic Telescope (LAMOST) is a Chinese national scientific research facility operated by the National Astronomical Observatories, Chinese Academy of Sciences. It is a special reflecting Schmidt telescope with 4000 fibers in a field of view of 20 deg$^2$. The data of this release are available via http://dr7.lamost.org/v1.3/.    

LAMOST DR7 includes a total number of 10,640,255 low resolution spectra covering the whole optical wavelength range of 3690 -- 9100Å at a spectral resolution of about 1800. The LAMOST Stellar Parameter Pipeline (LASP; Wu et al. 2011) has been used to determine the basic stellar parameters including effective temperature  {\teff}, surface gravity {\logg}, metallicity {\feh}, and radial velocity $V_{\rm r}$ for late A and FGK stars. 
Their typical errors compared with SDSS DR9  are $-91 \pm 111$\,K for {\teff}, $0.16 \pm 0.22$\, for {\logg}, and $0.04 \pm 0.15$\, for {\feh} (see Table 2 of Luo et al. 2015).
Repeated observations show that the internal errors of LAMOST parameters at SNR $>$ 20 are about 50 -- 100\,K for {\teff}, 0.05 -- 0.1\,dex for {\logg}, and 0.05 -- 0.1\,dex for {\feh} (See Figure\,3 of Niu et al. 2021a).

\subsection{SDSS Data Release 12} 
SDSS Data Release 12 (DR12; Alam et al. 2015) is the final data release of the SDSS-III, containing all SDSS observations through July 2014. 
The SDSS stellar spectra are processed through the SEGUE Stellar Parameter Pipeline (Lee et al. 2008a), and three primary stellar parameters ({\teff},{\logg}, and {\feh}) are obtained for most stars over the range of 4000 – 10,000 K and with spectral S/N ratios higher than 10.

\subsection{ {\it Gaia} Early Data Release 3} 
{\it Gaia} is an European Space Agency (ESA) mission aiming to map over one billion stars in the Milky Way in three dimensions.
Its EDR3 (Gaia Collaboration et al. 2021) is based on the first 34 months of the mission, including approximately 1.8 billion sources with precise astrometric and photometric information in 
three bands ($G$, $BP$, and $RP$). 
Modest calibration errors up to 10 mmag with G magnitude are found for stars of 10 $<$ G $<$ 19 (Yang et al. 2021). Hence we apply the magnitude corrections of $BP$ and $RP$ (see Table\,1 of Yang et al. 2021) in this work.

\section{The Recalibration of Stripe 82} 

\begin{figure*}
\centering
\includegraphics[width=160mm]{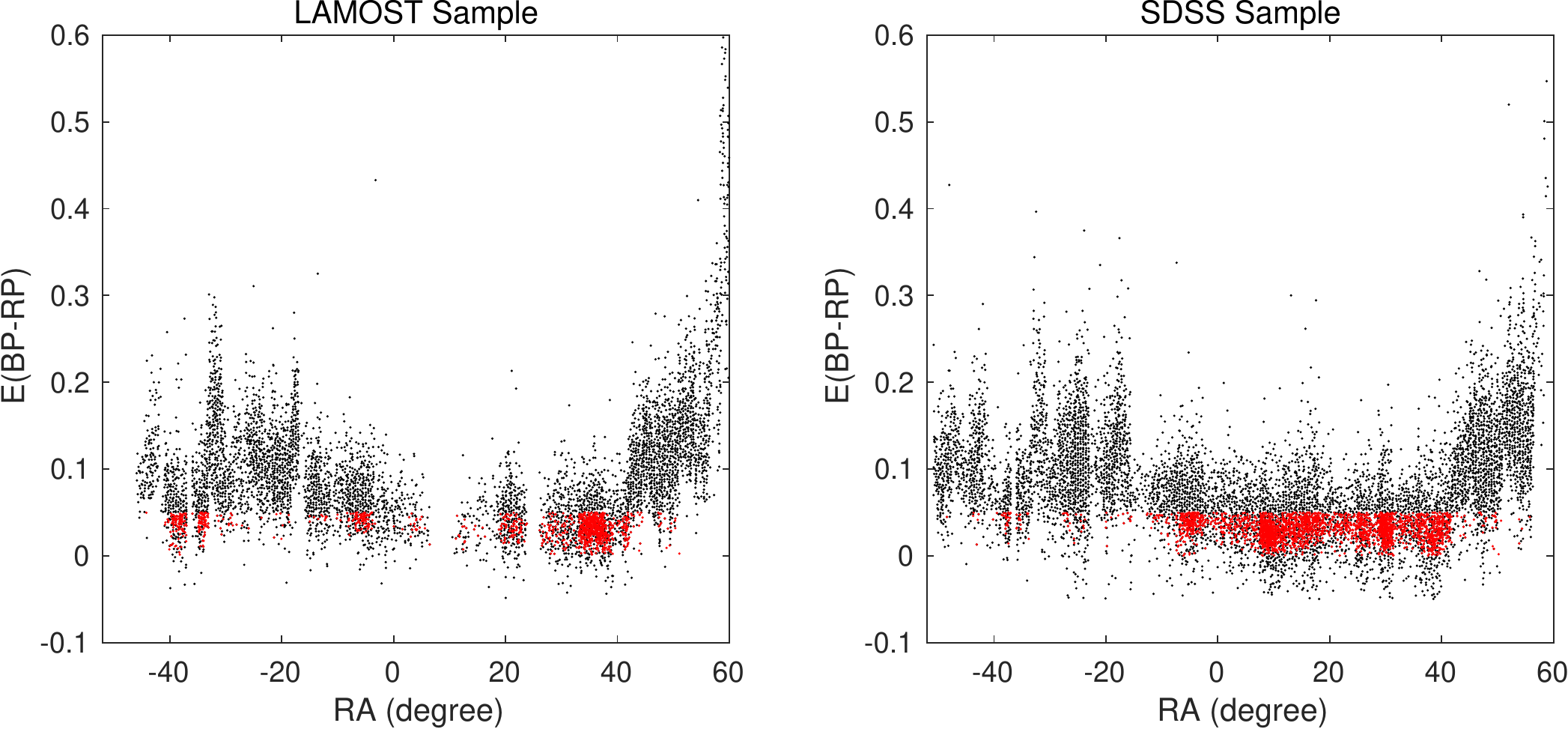}
\caption{Distributions in the R.A. -- \ebprp~plane for the LAMOST (left) and  SDSS (right) target stars.  The LAMOST and SDSS reference stars are also plotted in red dots.} 
\label{}
\end{figure*}

\begin{figure*}
\centering
\includegraphics[width=160mm]{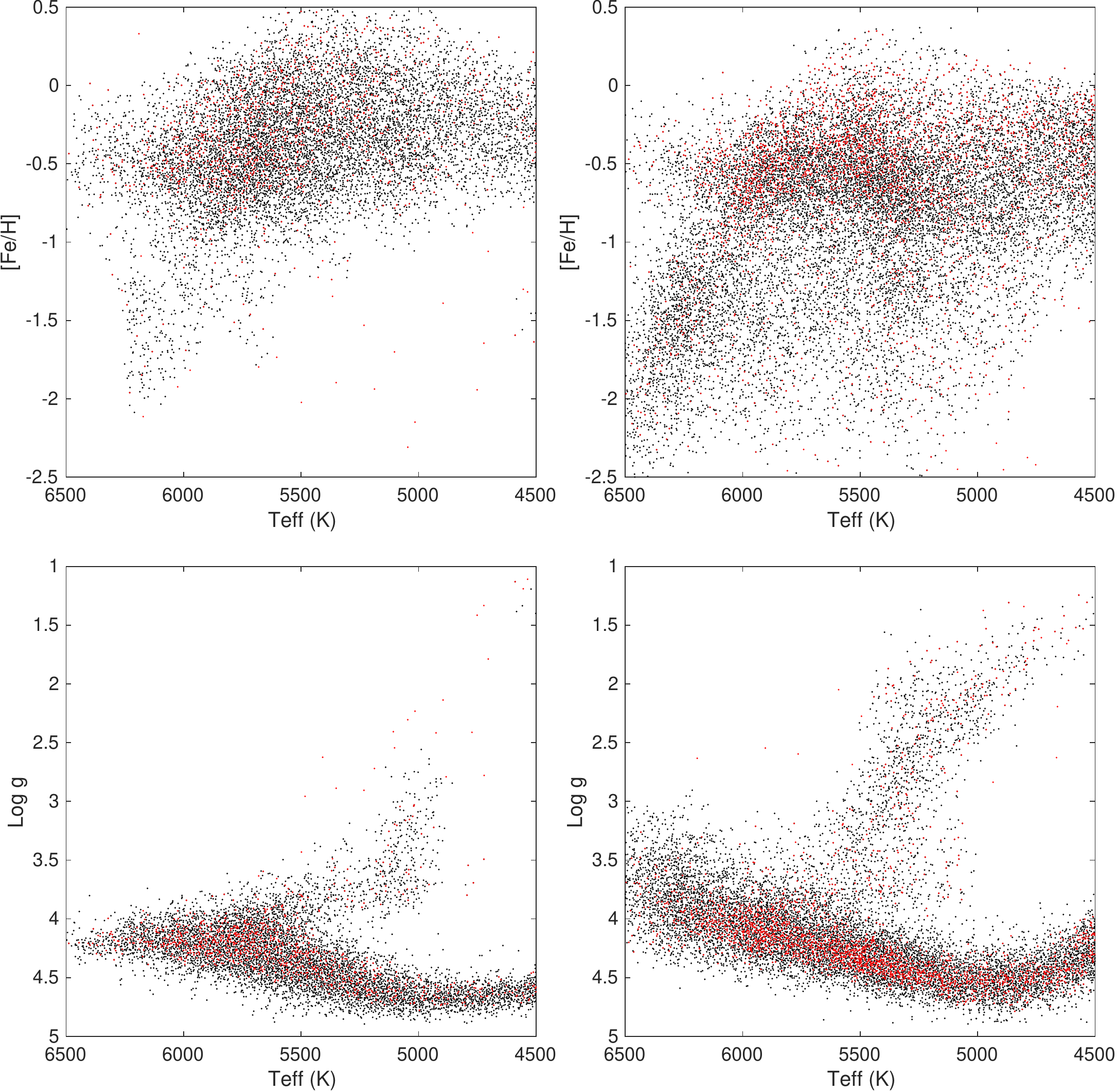}
\caption{Distributions in the planes of {\teff}~-- {\feh} (top) and {\teff}~-- {\logg} (bottom) for the LAMOST (left) and  SDSS (right) target stars.  The reference stars are plotted red dots.} 

\label{}
\end{figure*}

\subsection{Data Selection}
There are over 20,000 common sources between I07, LAMOST DR7, and {\it Gaia} EDR3,  and over 28,000 common sources between I07, SDSS DR12, and {\it Gaia} EDR3,  adopting a matching radius of 1 arcsec. 
The two sets of common sources are called respectively the LAMOST sample and the SDSS sample hereafter. 
We apply the SCR method using the two sets of data independently at the beginning. In the end, their results are combined, and compared with each other.

Reddening values of stars are required in the SCR method. In Yuan et al. (2015a), the dust reddening map of Schlegel et al. (1998; SFD) was used. However, the SFD map shows spatially dependent systematic 
errors (e.g., Sun et al., submitted). Therefore, we use \ebprp~values determined with the star pair technique (Yuan et al. 2013; Ruoyi \& Haibo 2020) in this work. The \ebprp~values of the LAMOST and SDSS samples are determined separately.

We firstly select reference samples that with SNR$_g$ (signal-to-noise ratio in the $g$ band) higher than 40 and having low \ebprp~values between 0 $-$ 0.05.  The reference samples are used to define the stellar intrinsic colors as a function of stellar atmospheric parameters with linear interpolation. 
With the above criteria, we have 1,182 and 2,849 reference stars from the LAMOST and SDSS samples, respectively.

Then we select target samples with criteria that SNR$_g$ higher than 15, \ebprp~greater than $-$0.05 and effective temperature {\teff} between 4,500\,K and 6,500\,K. 
The cut on {\teff} is mainly for robust fitting of temperature-dependent reddening coefficients 
(e.g., Niu et al. 2021a,b). 
We also require that each target star should have at least 4 reference stars whose 
\teff, \logg, and \feh~values differ by less than 100\,K, 0.5\,dex, and 0.3\,dex, respectively.
With the above criteria, we have 11,477 and 18,239 target stars 
selected from the LAMOST and SDSS samples, respectively. 
These stars are called the target samples hereafter. 
Properties of the target and reference samples are plotted in Figure\,1 and Figure\,2. The reference stars appropriately sample the range of stellar parameters seen in the full data set, as seen in Figure\,2.

\subsection{Reddening corrections and intrinsic colors} 
To perform photometric calibration for the $u$, $g$, $r$, $i$, and $z$ bands with the SCR method, 
five colors $BP-u$, $BP-g$, $RP-r$, $RP-i$, and $RP-z$ are adopted respectively, 
where $BP$ and $RP$ are from {\it Gaia} EDR3. 

Reddening coefficients of these colors with respect to \ebprp~ are required to obtain the intrinsic colors of the target samples. 
The reddening coefficients are derived from the following steps.
Firstly, with an initial set of reddening coefficients,  the reference stars are dereddened. 
Due to the very broad $BP$ and $RP$ passbands, temperature-dependent reddening coefficients are adopted here. For simplicity, we assume that the reddening coefficients are linear functions of effective temperature.
Then we estimate color excess values of the target stars, with their intrinsic colors determined spectroscopically from the dereddened reference sample by the star-pair technique. 
Finally, a new set of reddening coefficients are derived by the least-squares fitting method.
A 2.5-$\sigma$ clipping is performed in the fitting. 
The above process is iterated, till the new reddening coefficients are consistent with previous ones. 
Note that the reddening coefficients are derived using only the LAMOST stars. The yielded coefficients are used for the SDSS stars as well. The results of temperature-dependent reddening coefficients for the five colors are listed
in Table\,1 and displayed in Figures\,3--4.

\begin{table}
\centering
\caption{Temperature-dependent reddening coefficients with respect to \ebprp.}
\label{} \small 
\begin{tabular}{ccc} \hline\hline
Color & $a_1$ & $a_0$\\ \hline
$BP - u$ & $+1.333 \times 10^{-4}$ & $-1.850$ \\
$BP - g$ & $+0.592 \times 10^{-4}$ & $-0.641$ \\
$RP - r$ & $+0.755 \times 10^{-4}$ & $-0.952$ \\
$RP - i$ & $+0.272 \times 10^{-4}$ & $-0.217$ \\
$RP - z$ & $-0.098 \times 10^{-4}$ & $+0.384$ \\
\hline
$R = a_1 \times \teff + a_0$
\end{tabular}
\end{table}

Note that the spatial variations of reddening coefficients across the Stripe 82 are neglected, to avoid possible degeneracy between zero-point offsets and reddening coefficients. Fortunately, the Stripe 82 is located at high Galactic latitudes and has small values of reddening. Thus, the impacts of possible spatially varying reddening coefficients on our results are small.

\begin{figure}
\includegraphics[width=90mm]{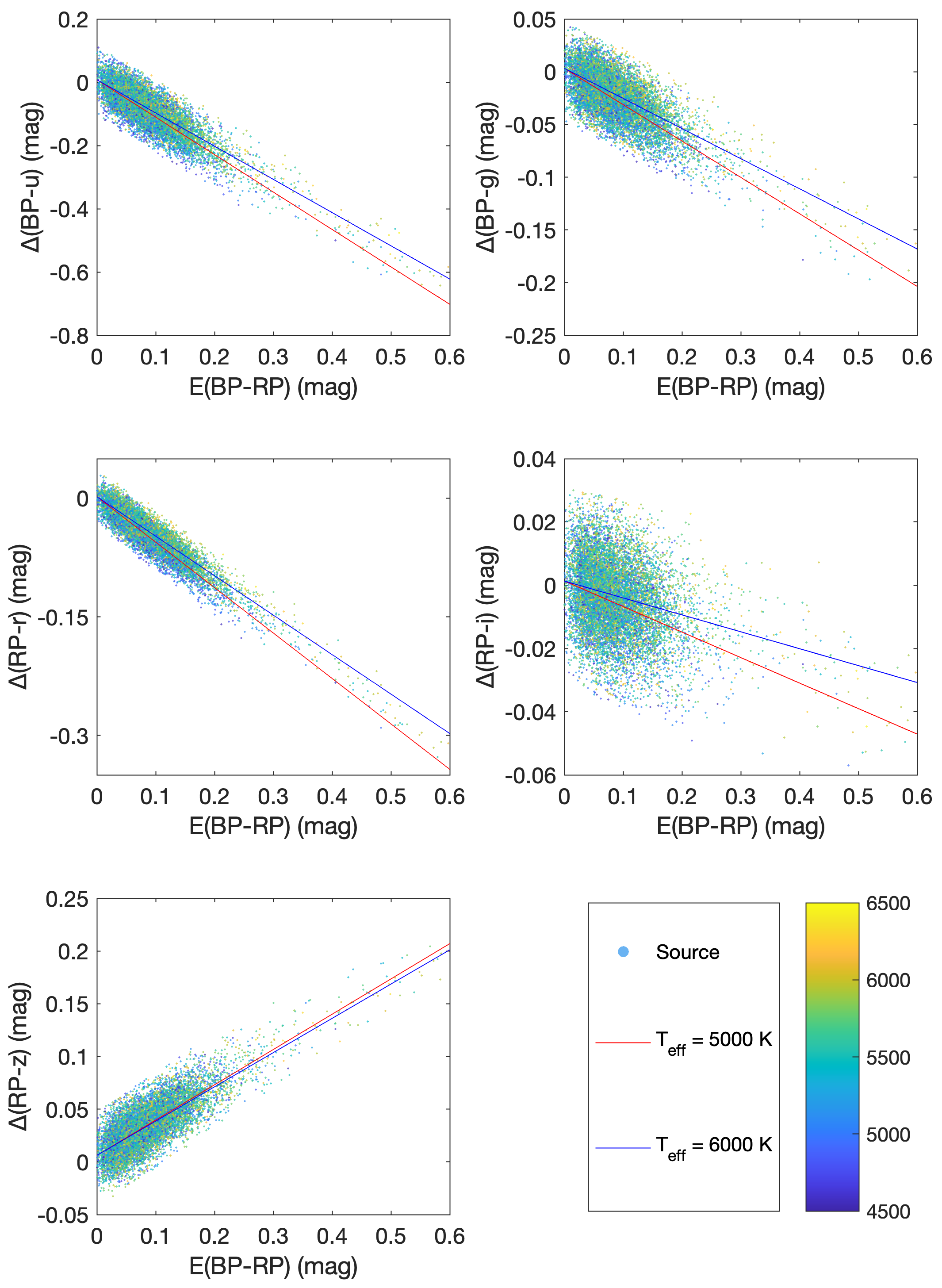}
\caption{Reddening values in $BP-u$, $BP-g$, $RP-r$, $RP-i$, and $RP-z$ colors plotted against \ebprp~ for the LAMOST sample with 2.5-$\sigma$ clipping performed. The red and blue lines mark the reddening coefficients at 5,000\,K and 6,000\,K, respectively. Note that lines are not forced to go through the origin.}
\label{}
\end{figure}

\begin{figure}
\includegraphics[width=90mm]{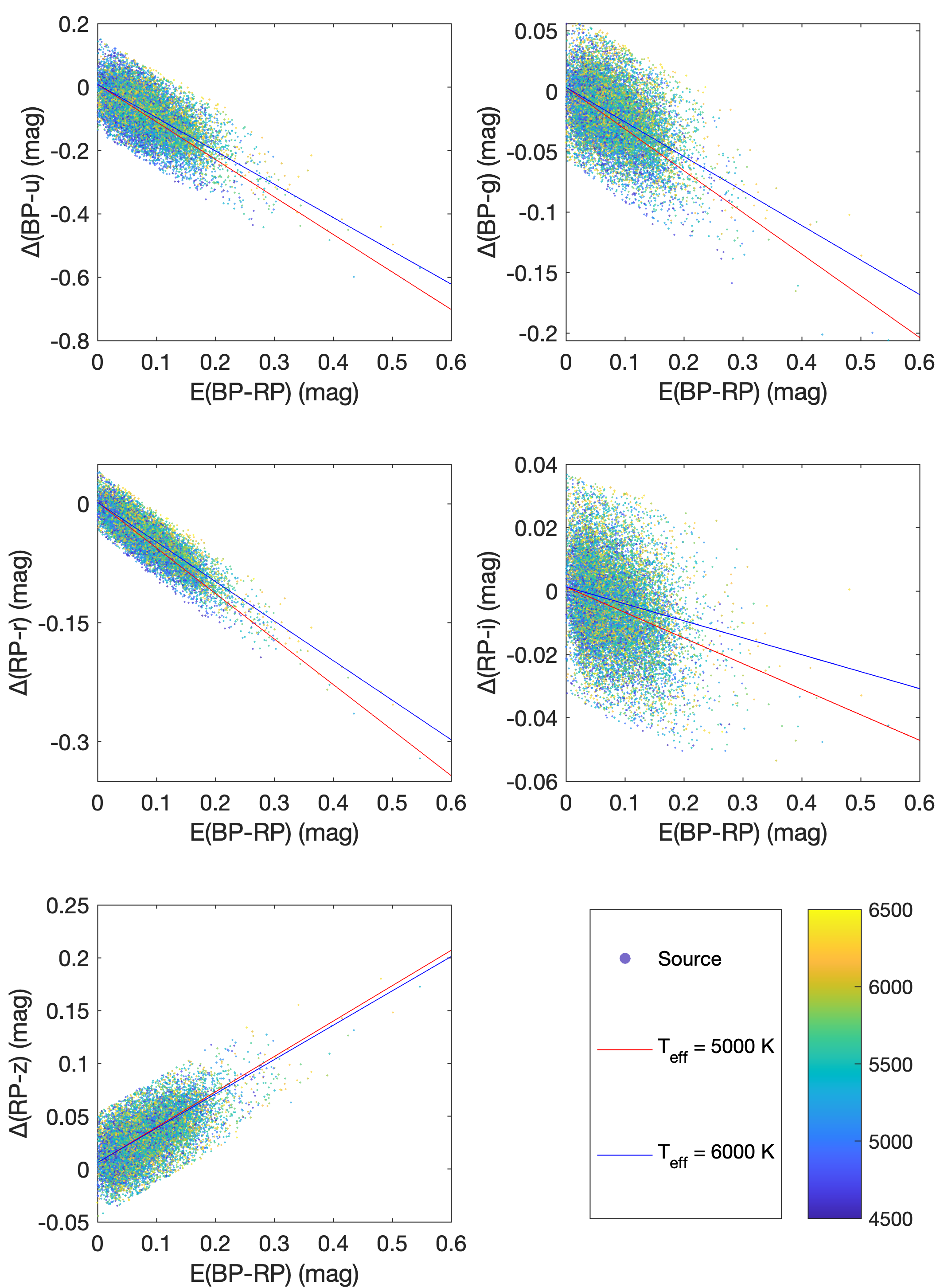}
\caption{Same to Figure\,3 but for the SDSS sample.}
\label{}
\end{figure}

In the above process, spectroscopy-based intrinsic colors of the target stars are also obtained. 
To test the precision of the predicted intrinsic colors, the color fitting residuals are plotted 
in Figures\,5--6 for the LAMOST and SDSS target samples, respectively. For the LAMOST target sample, 
the typical errors are 0.039, 0.017, 0.012, 0.011, and 0.015 mag respectively for the predicted intrinsic colors of $BP-u$, $BP-g$, $RP-r$, $RP-i$, and $RP-z$. The residuals show no obvious trends with respect to the atmospheric parameters.  
For the SDSS target sample, the typical errors are slightly larger, as the SDSS target stars are fainter
and suffer larger photometric errors.

\begin{figure*}
\includegraphics[width=180mm]{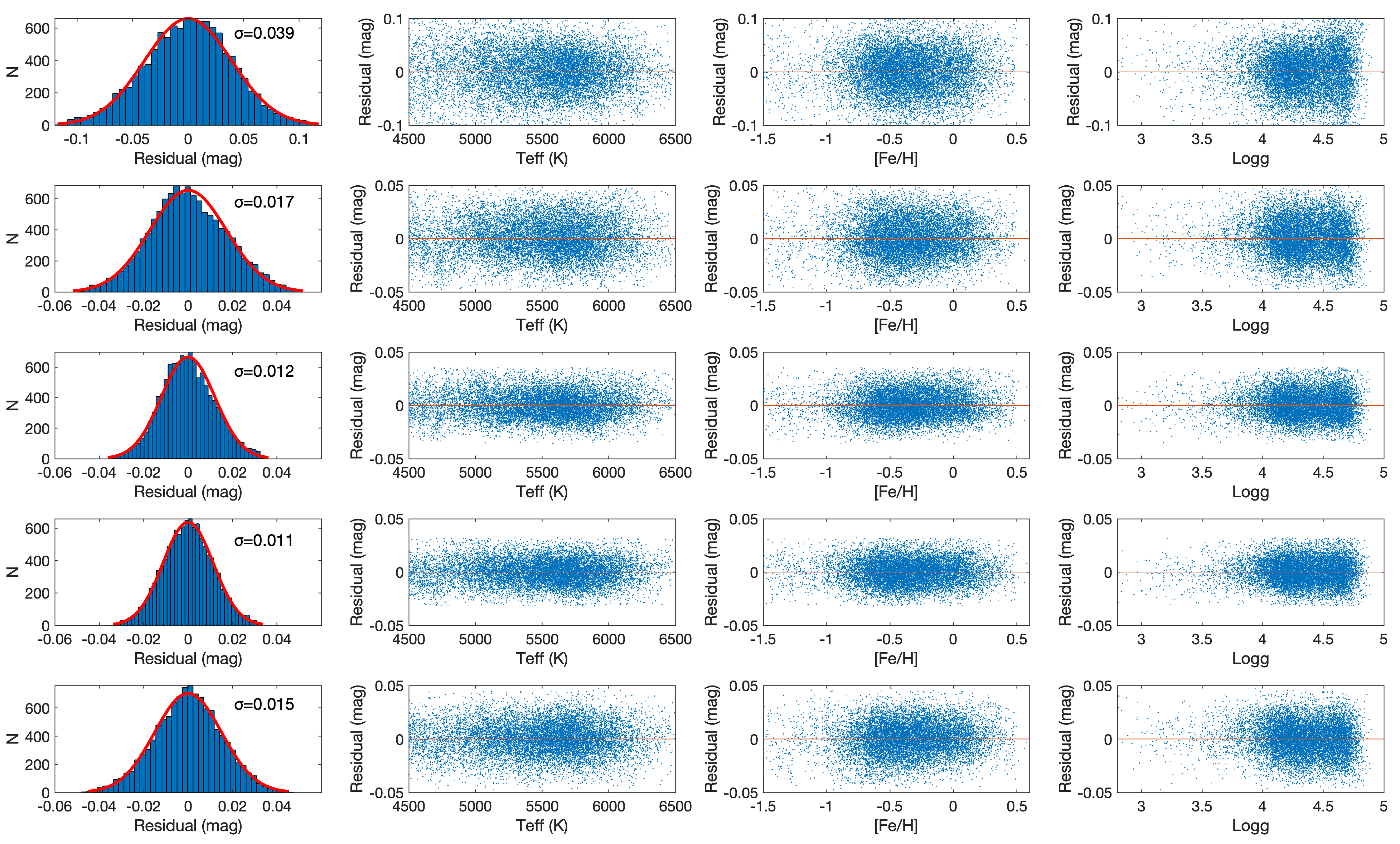}
\centering

\caption{Precision of intrinsic colors for the LAMOST target sample. From top to bottom are $BP-u$, $BP-g$, $RP-r$, $RP-i$, and $RP-z$ colors, respectively. 
The 1st column shows the residual distributions. 
The Gaussian fitting results are over-plotted in red, with the $\sigma$ values marked. The 2nd, 3rd, and 4th columns plot residuals against \teff, \feh, and \logg, respectively. }
\label{}
\end{figure*}

\begin{figure*}
\includegraphics[width=180mm]{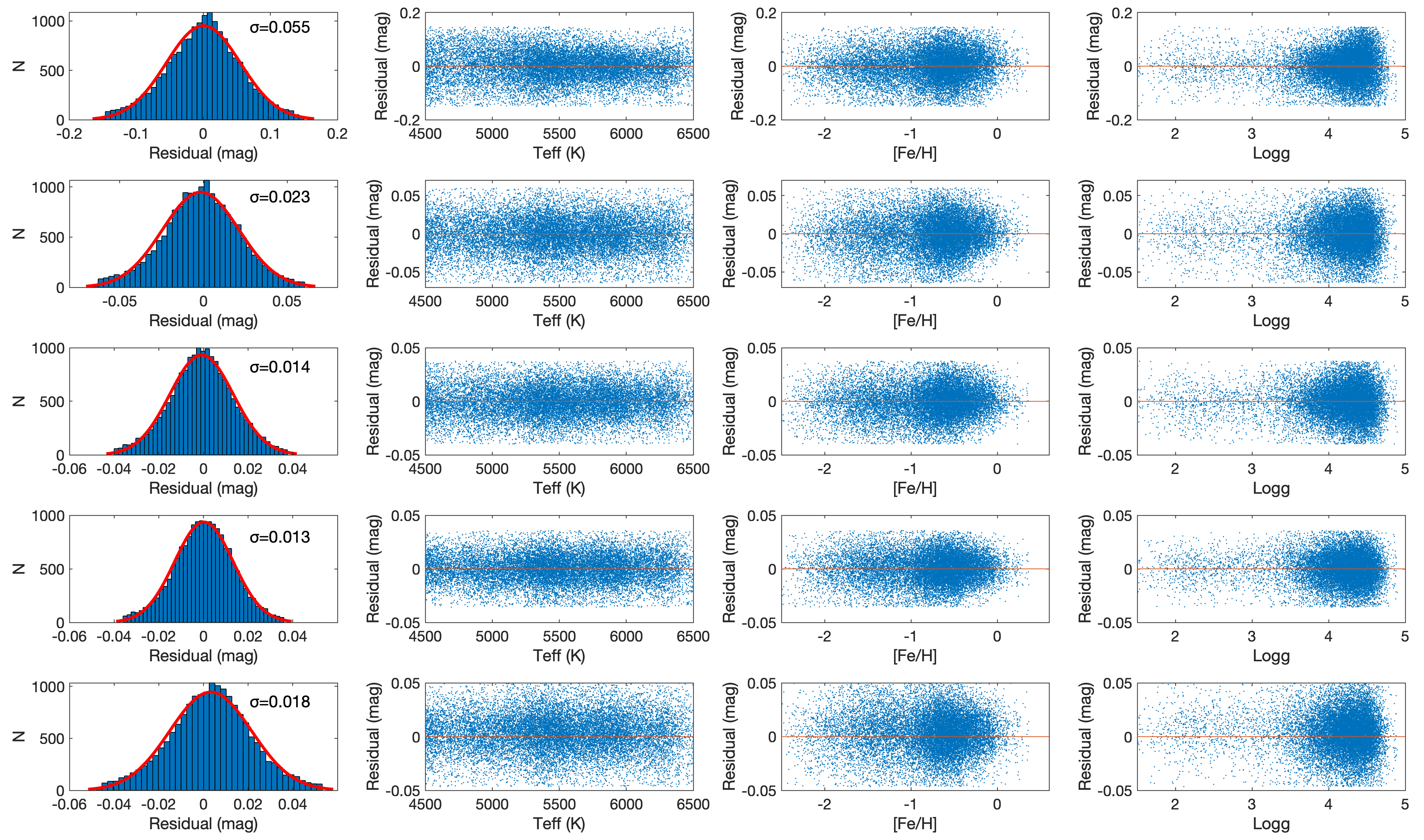}
\centering
\caption{Same to Figure$\,$5 but for the SDSS target sample. }
\label{}
\end{figure*}

\subsection{From Color Corrections to Photometric Corrections}
With the reddening coefficients and the intrinsic colors, the observed colors of the LAMOST and SDSS target samples are predicted. The predicted magnitudes in the $u$, $g$, $r$, $i$, and $z$ bands are then derived 
using the {\it Gaia} EDR3's $BP$ or $RP$ magnitudes. Then, the magnitude offsets in each SDSS band can be  determined by comparing the predicted magnitudes and the original magnitudes in I07, as shown in the next subsection.

\subsection{Decouple Spatial Variations of Magnitude Offsets in R.A. $\delta_m^{ext}({\rm R.A.})$ and in Dec. $\delta_m^{ff}({\rm Dec.})$}

Similar to color calibration (I07; Yuan et al. 2015a), the true magnitude of an object, $m_{\rm true}$, can be expressed as: 
\begin{equation} 
m_{true} = m_{cat} + \delta_m^{ext}({\rm R.A.}) + \delta_m^{ff}({\rm Dec.}),
\end{equation}
where $m_{\rm cat}$ is from the I07 catalog,
$\delta_m^{ext}({\rm R.A.})$ is dominated by fast variations of the atmospheric extinction, 
and $\delta_m^{ff}({\rm Dec.})$ is dominated by errors in the flat-field correction. 
The R.A. and Dec. dependence of the dominant errors comes from the drift scan direction for 
the Stripe\,82, which was along the Equator.
Unlike in Yuan et al. (2015a), we obtain $\delta_m^{ext}({\rm R.A.})$ and  $\delta_m^{ff}({\rm Dec.}$) 
simultaneously in this work.
Note that the LAMOST and SDSS target samples are combined in this step.

Stripe 82 consists of two drift scan strips: the north strip and the south strip. 
Each strip is scanned by six columns of five rows of CCDs with gaps between the columns. 
The geometry of the SDSS imaging can be found in Figure\,1 of Padmanabhan et al. (2008).
We decouple spatial variations of magnitude offsets in the two strips independently. 
For each strip, we uniformly divide the strip into 111 bins in R.A. and 49 bins in Dec.. 
The bin width in R.A. is 1 degree, and the bin width in Dec. is 1/8 of the CCD width. 
The magnitude offsets in each band of I07 are displayed in Figures\,7--8.

\begin{figure*}
\includegraphics[width=180mm]{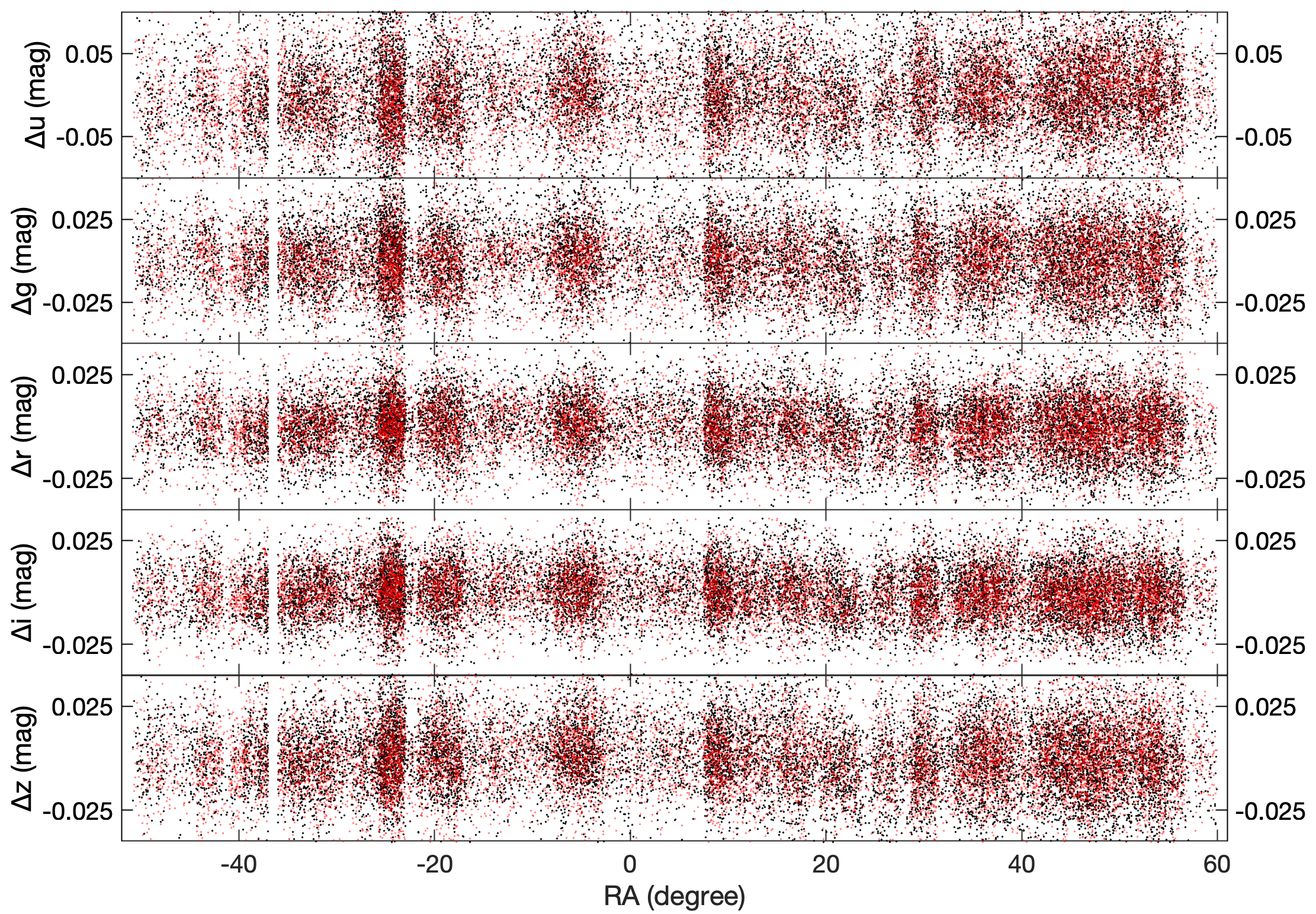}
\centering
\caption{Magnitude offsets in the $u$, $g$, $r$, $i$, and $z$ bands as a function of R.A. for the LAMOST and SDSS target samples. Stars in the north and south strips are displayed in red and black, respectively.}
\label{}
\end{figure*}

\begin{figure*}
\includegraphics[width=180mm]{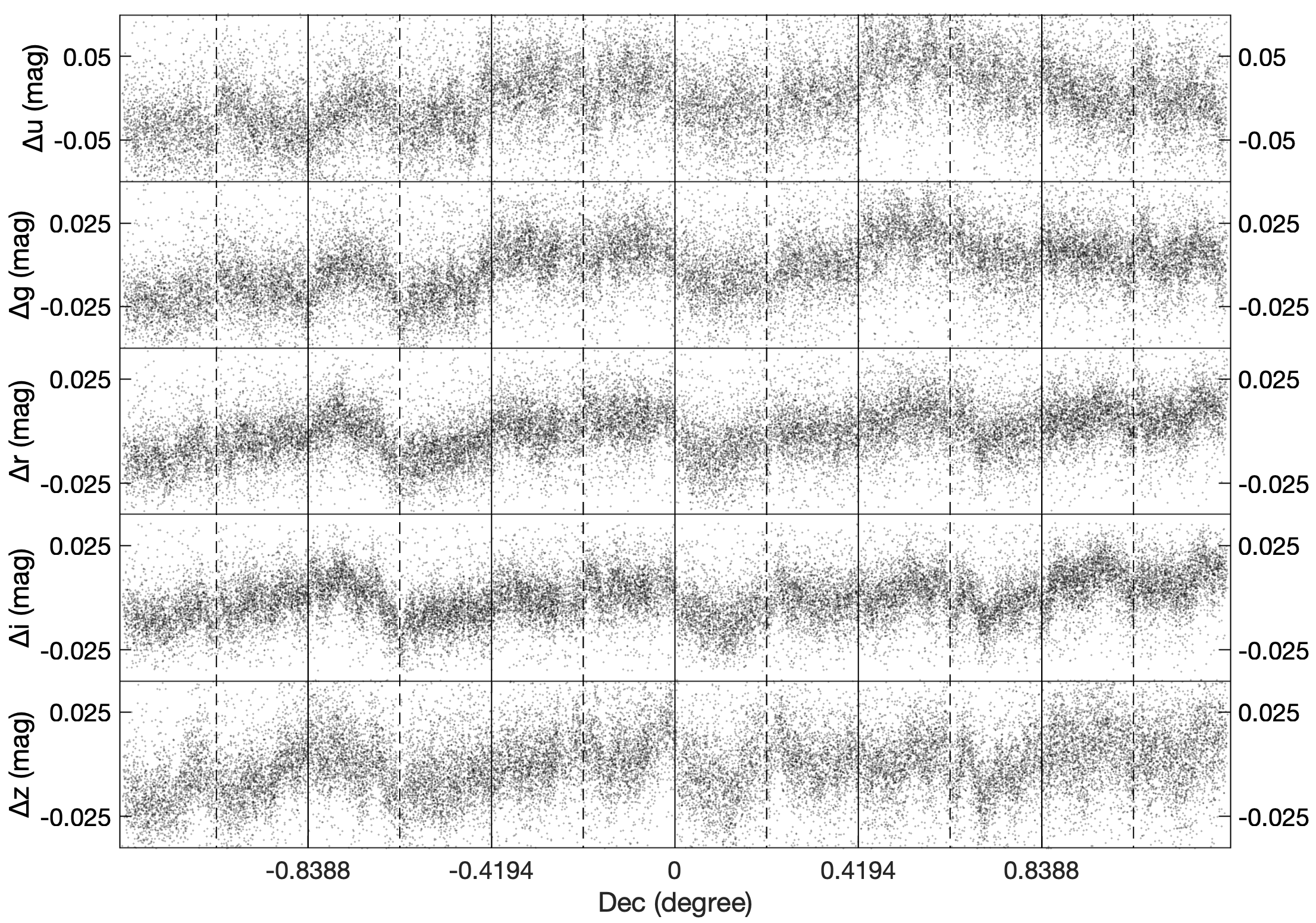}
\centering
\caption{Magnitude offsets in the $u$, $g$, $r$, $i$, and $z$ bands as a function of Dec. for the LAMOST and SDSS target samples. The vertical solid lines mark the approximate boundaries between the different CCD columns. The vertical dashed lines mark the approximate boundaries between the two strips of Stripe 82.}
\label{}
\end{figure*}

To decouple the spatial variations of magnitude offsets, we assume that each grid has independent $\delta_m^{ext}({\rm RA})$ and $\delta_m^{ff}({\rm Dec})$. 
According to the above equation, we apply it to the LAMOST and SDSS target samples. With the least-squares method, we obtain 160 free parameters (111 in $\delta_m^{ext}({\rm R.A.})$ and 49 in $\delta_m^{ff}({\rm Dec.})$) simultaneously for each strip. 
To avoid the degeneracy between $\delta_m^{ext}({\rm R.A.})$ and $\delta_m^{ff}({\rm Dec.})$, 
we require that the mean value of $\delta_m^{ff}({\rm Dec.})$ equals zero for each band. 
The results are displayed in Figures\,9--10, and listed in Tables\,A1--A3 in the Appendix. The entries in the tables should be added to cataloged photometry, 
as shown in Equation (1).
Histogram distributions of the combined $\delta_m^{ext}({\rm R.A.})$ and $\delta_m^{ff}({\rm Dec.})$ values 
in the Stripe 82 region are displayed in Figure\,11. The calibration errors of I07 are estimated to be 
0.025, 0.011, 0.007, 0.007, and 0.008 mag for the $u$, $g$, $r$, $i$, and $z$ bands, respectively.  

\begin{figure*}
\includegraphics[width=180mm]{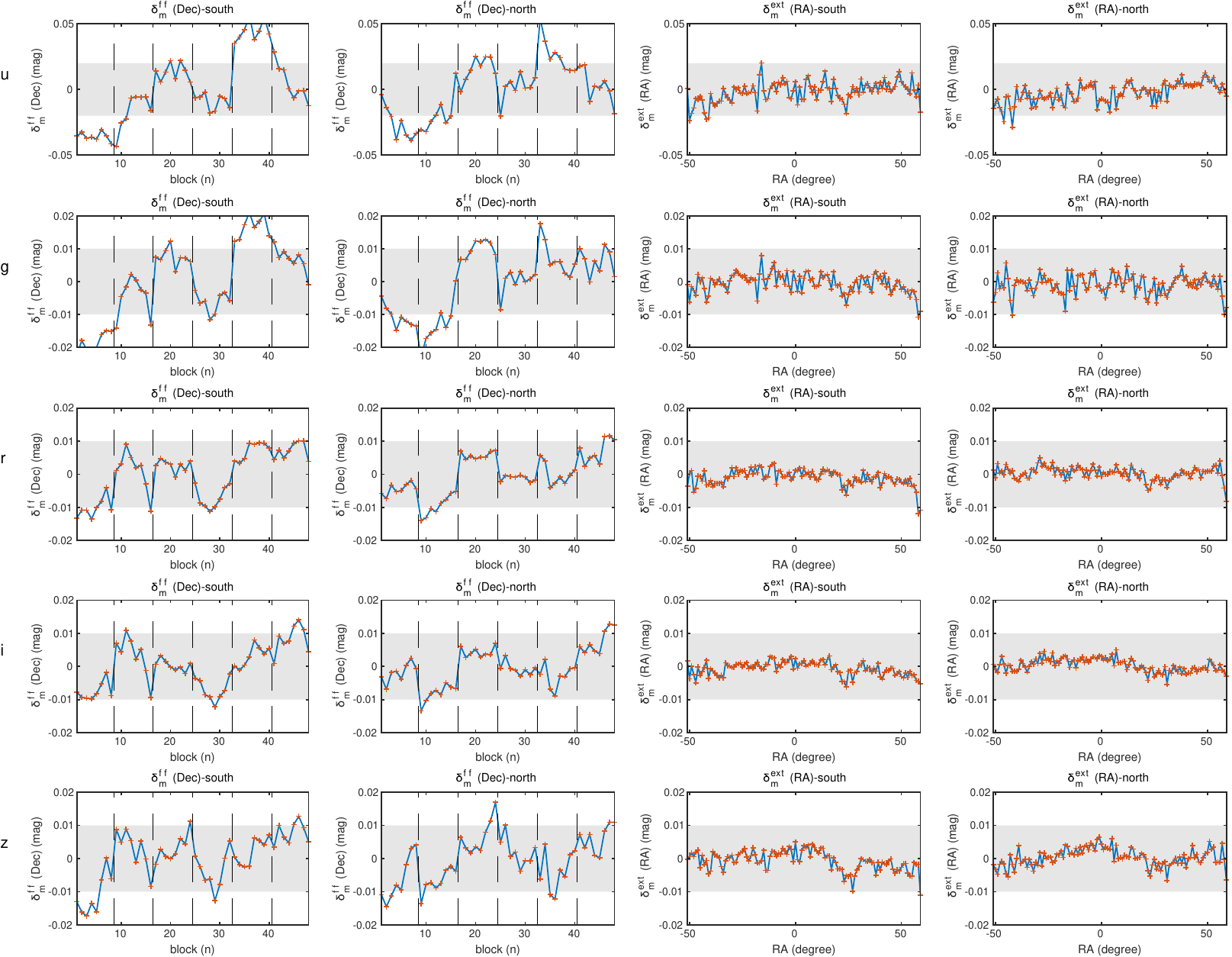}
\caption{$\delta_m^{ext}({\rm R.A.})$ and $\delta_m^{ff}({\rm Dec.})$ for the $u$, $g$, $r$, $i$, and $z$ bands. The 1st and 2nd columns plot $\delta_m^{ff}({\rm Dec.})$ as a function of Dec. for the south and north strips, respectively. The vertical dashed lines mark the approximate boundaries between the different camera CCD columns.
The 3rd and 4th columns plot $\delta_m^{ext}({\rm R.A.})$ as a function of R.A. for the south and north strips, respectively. The shaded regions are of $\pm 0.02$\,mag in the $u$ band and $\pm 0.01$\,mag in the $griz$ bands.}
\label{}
\end{figure*}

\begin{figure}
\includegraphics[width=90mm]{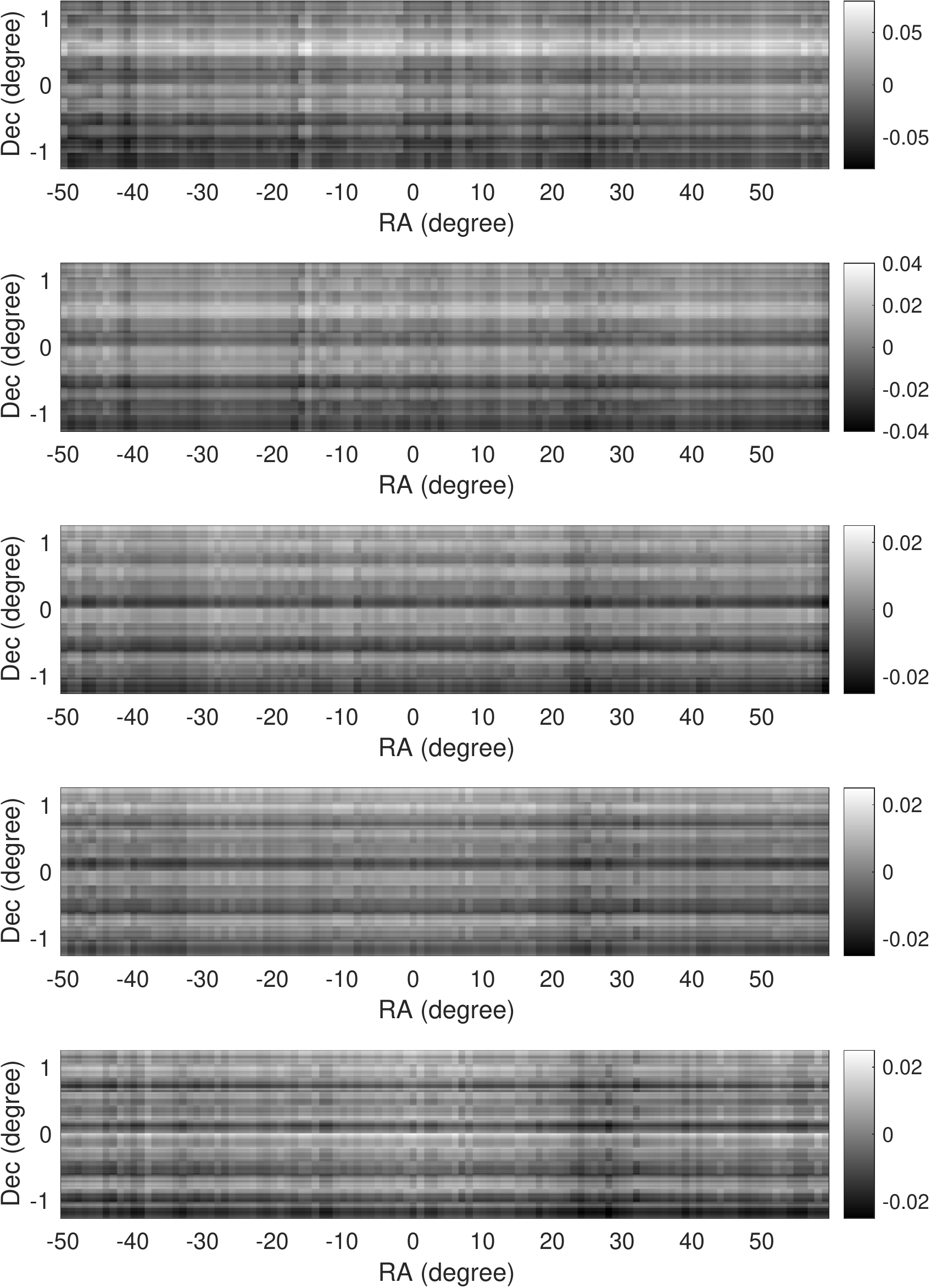}
\caption{Spatial variations of magnitude offsets ($\delta_m^{ext}({\rm R.A.})$ + $\delta_m^{ff}({\rm Dec.})$) in the $u$, $g$, $r$, $i$, and $z$ bands. Colorbars are to the right of each panel.}
\label{}
\end{figure}

\begin{figure}
\includegraphics[width=90mm]{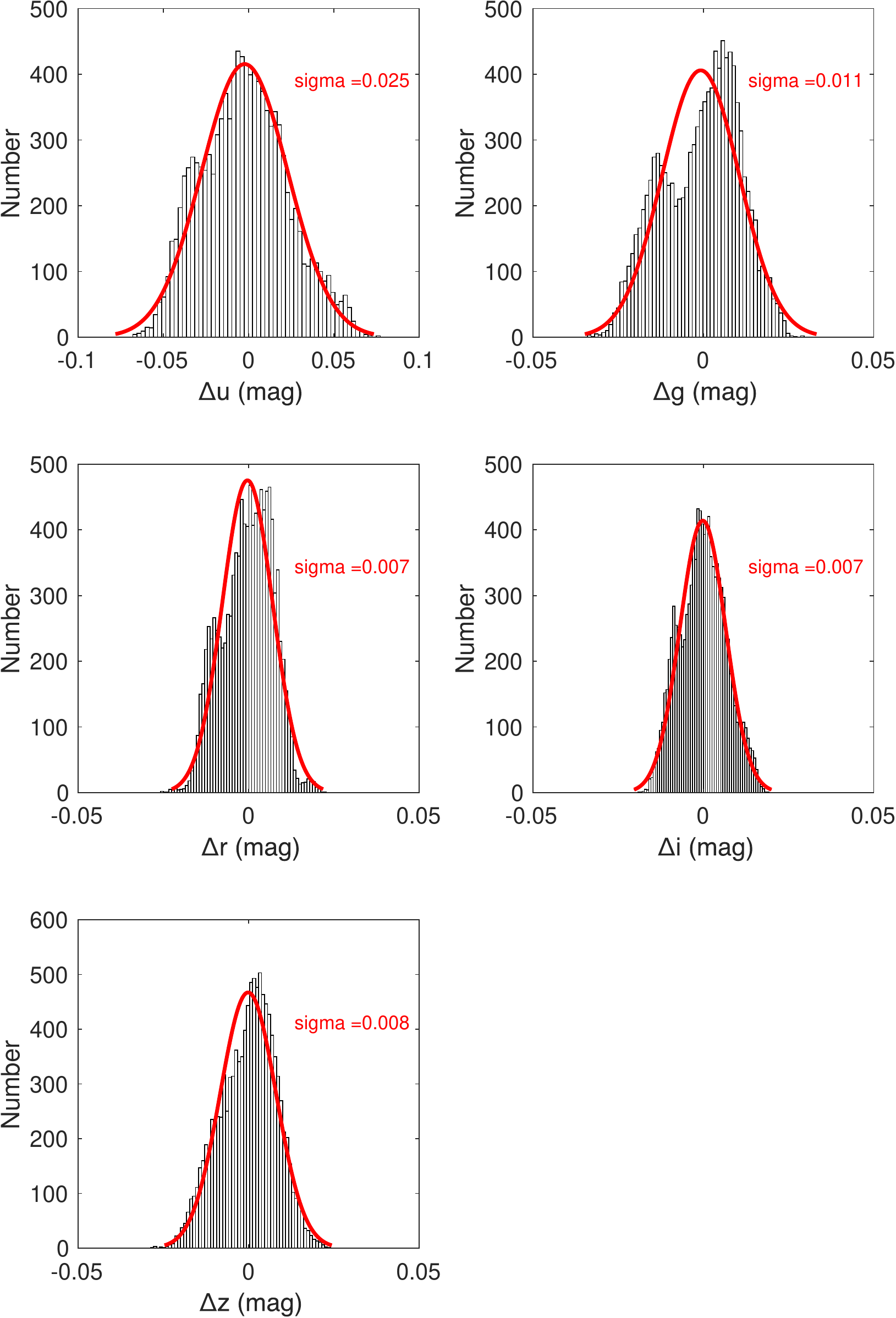}
\caption{Histogram distributions of magnitude offsets ($\delta_m^{ext}({\rm R.A.})$ + $\delta_m^{ff}({\rm Dec.})$) in the $u$, $g$, $r$, $i$, and $z$ bands. The solid red lines are Gaussian fitting results. The $\sigma$ values are 
marked.}
\label{}
\end{figure}

After corrections of the $\delta_m^{ext}({\rm R.A.})$ and $\delta_m^{ff}({\rm Dec.})$, 
magnitude offsets of the target samples are displayed in Figure\,12.  
No more offsets are found, as expected. 
Histogram distributions of the magnitude offsets in the $u$, $g$, $r$, $i$, and $z$ bands before and after corrections are displayed in Figure\,13.  
Unsurprisingly, the dispersion values have decreased significantly, by 0.032, 0.014, 0.0087, 0.0081, and 0.010
mag in the $u$, $g$, $r$, $i$, and $z$ bands, respectively.
The decreasements are consistent with the calibration errors of I07. 

\begin{figure}
\includegraphics[width=90mm]{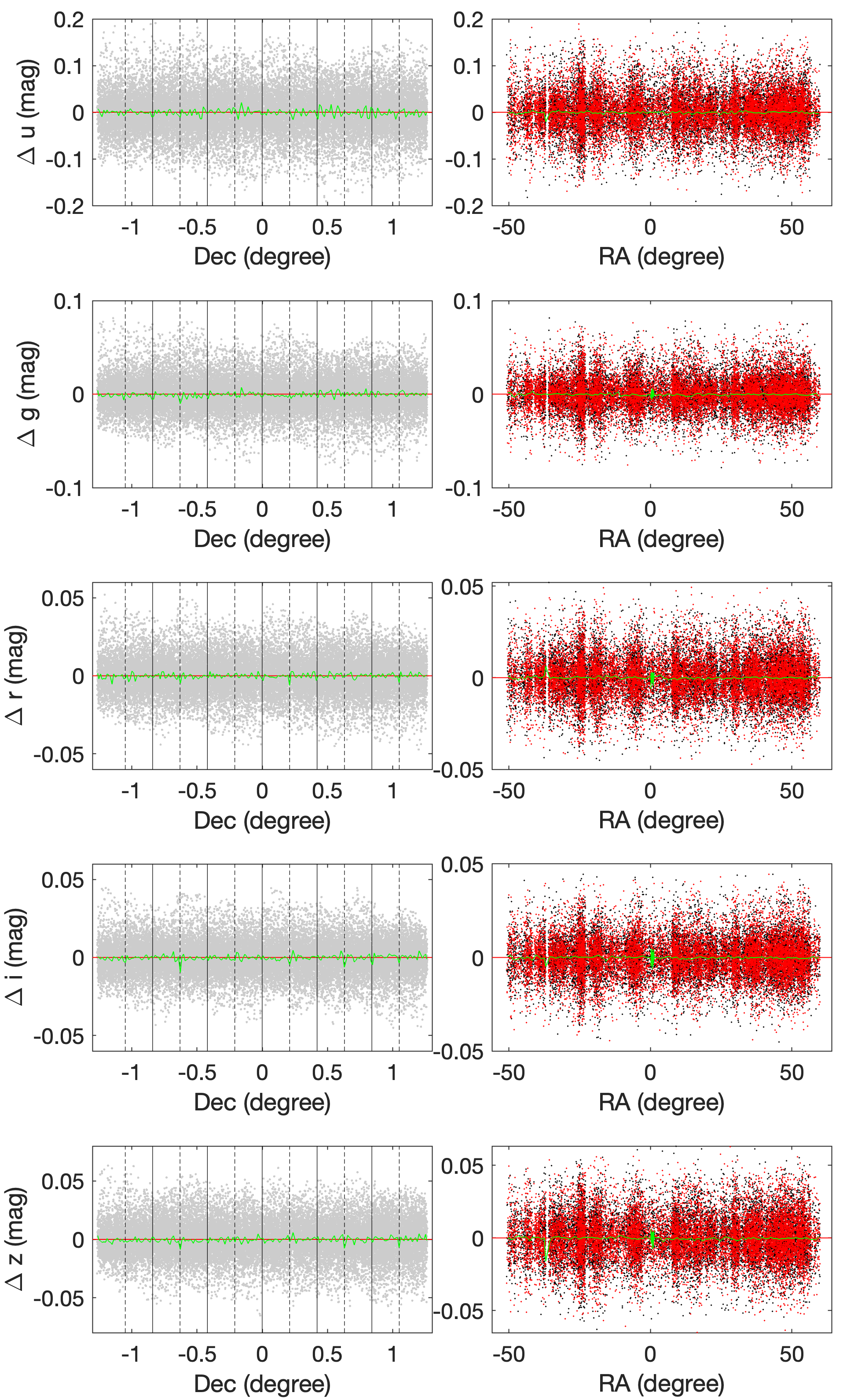}
\caption{Magnitude offsets in the $u$, $g$, $r$, $i$, and $z$ bands as functions of Dec. (left) and R.A. (right), after corrections of the $\delta_m^{ext}({\rm R.A.})$ and $\delta_m^{ff}({\rm Dec.})$. 
In the left column, the vertical solid lines mark the approximate boundaries between the different CCD columns, and the vertical dashed lines mark the approximate boundaries between the two strips.
In the right column, the red and black dots represent stars in the north and south strips, respectively. 
The green lines represent the median values.}
\label{}
\end{figure}

\begin{figure}
\includegraphics[width=90mm]{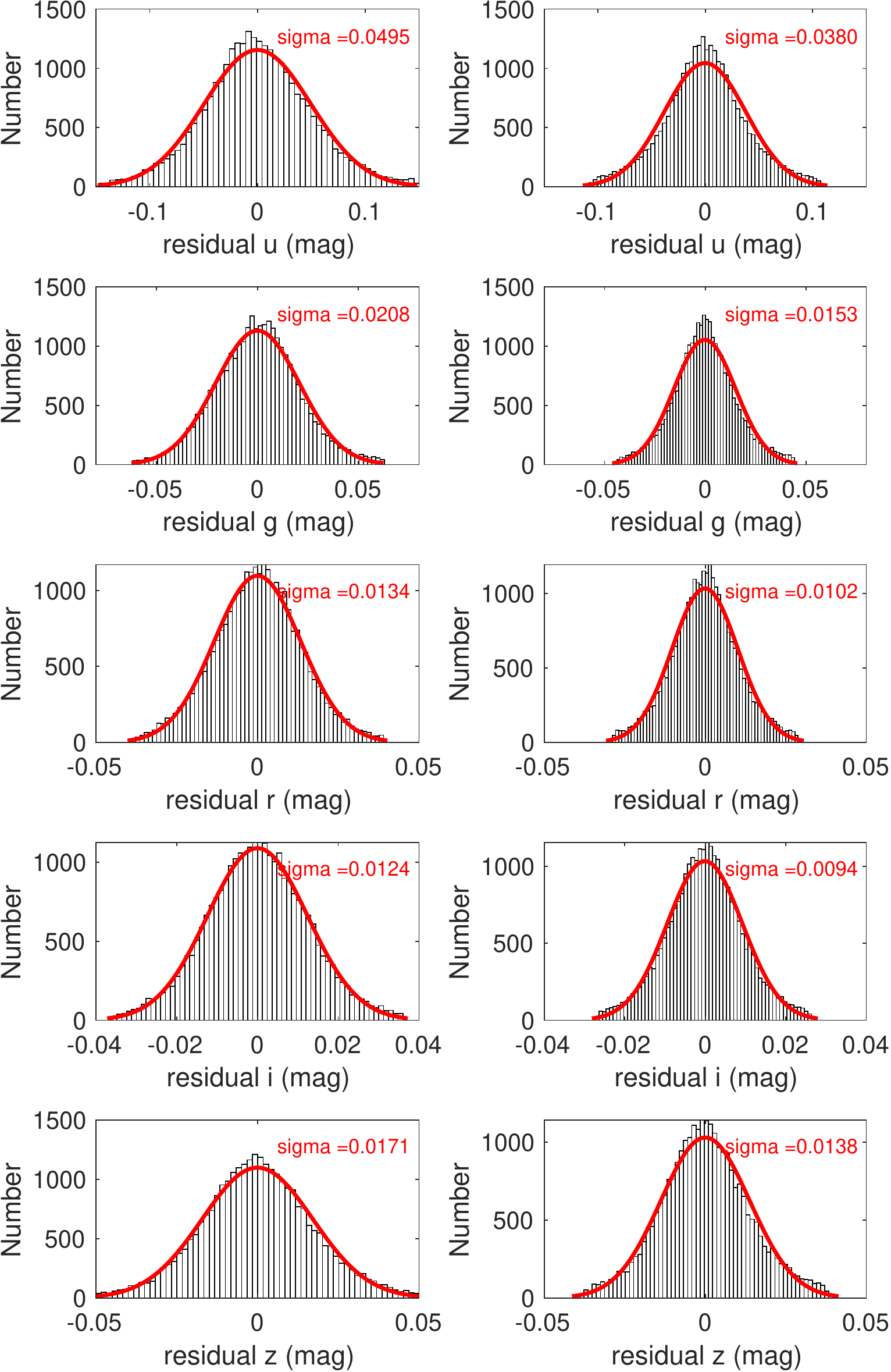}
\caption{Histogram distributions of magnitude offsets in the $u$, $g$, $r$, $i$, and $z$ bands before (left) and after (right) corrections of the $\delta_m^{ext}({\rm R.A.})$ and $\delta_m^{ff}({\rm Dec.})$. 
The Gaussian fitting curves are over-plotted, with the $\sigma$ values labeled.}
\label{}
\end{figure}

\subsection{Dependence on Magnitudes and Colors}
After corrections of the $\delta_m^{ext}({\rm R.A.})$ and $\delta_m^{ff}({\rm Dec.})$, 
we investigate the possible dependence of the remaining magnitude offsets on target magnitudes and colors.
The results are shown in Figures\,14--15.

Obvious variations are only found in three (camcol=2/3/6) of the six CCD columns for the $z$ band, 
confirming the results of Yuan et al. (2015a) that the variations are caused by the un-corrected non-linearity of the $z$ band detectors. 
We have performed a third-order polynomial fit to the observed variations of $z$ magnitude offsets as a function of $z$ magnitude for the aforementioned three CCD columns. The results are over-plotted in red in Figure\,14. The fit coefficients are listed in Table\,2 and valid for $14 < z < 18.5$.
The corrections should also be added to the cataloged photometry. For stars brighter than 14 or fainter than 18.5, 
the correction values at 14 or 18.5 are adopted, respectively.

No obvious variations are found for the  magnitude offsets as a function of $g-i$ color.
However, for red stars of $g-i > 1$, 
there seem to be weak color dependences for the $u$ magnitude offsets in 
the first two CCD columns. Further investigations are needed. 

\begin{table}
\centering
\caption{Fit coefficients for the $z$ magnitude offsets as a function of $z$ magnitude.}
\label{} \small 
\begin{tabular}{rrrrr} \hline\hline
 Camcol & $z^3$ & $z^2$ & $z$ & Constant term\\ \hline
 2 & $-0.0008145$ & $0.03857$ & $-0.6021$ & $3.098$\\
 3 & $-0.0005567$ & $0.02618$ & $-0.4073$ & $2.094$\\
 6 & $-0.0014565$ & $0.06873$ & $-1.0732$ & $5.543$\\
\hline
\end{tabular}
\end{table}

\begin{figure*}
\includegraphics[width=180mm]{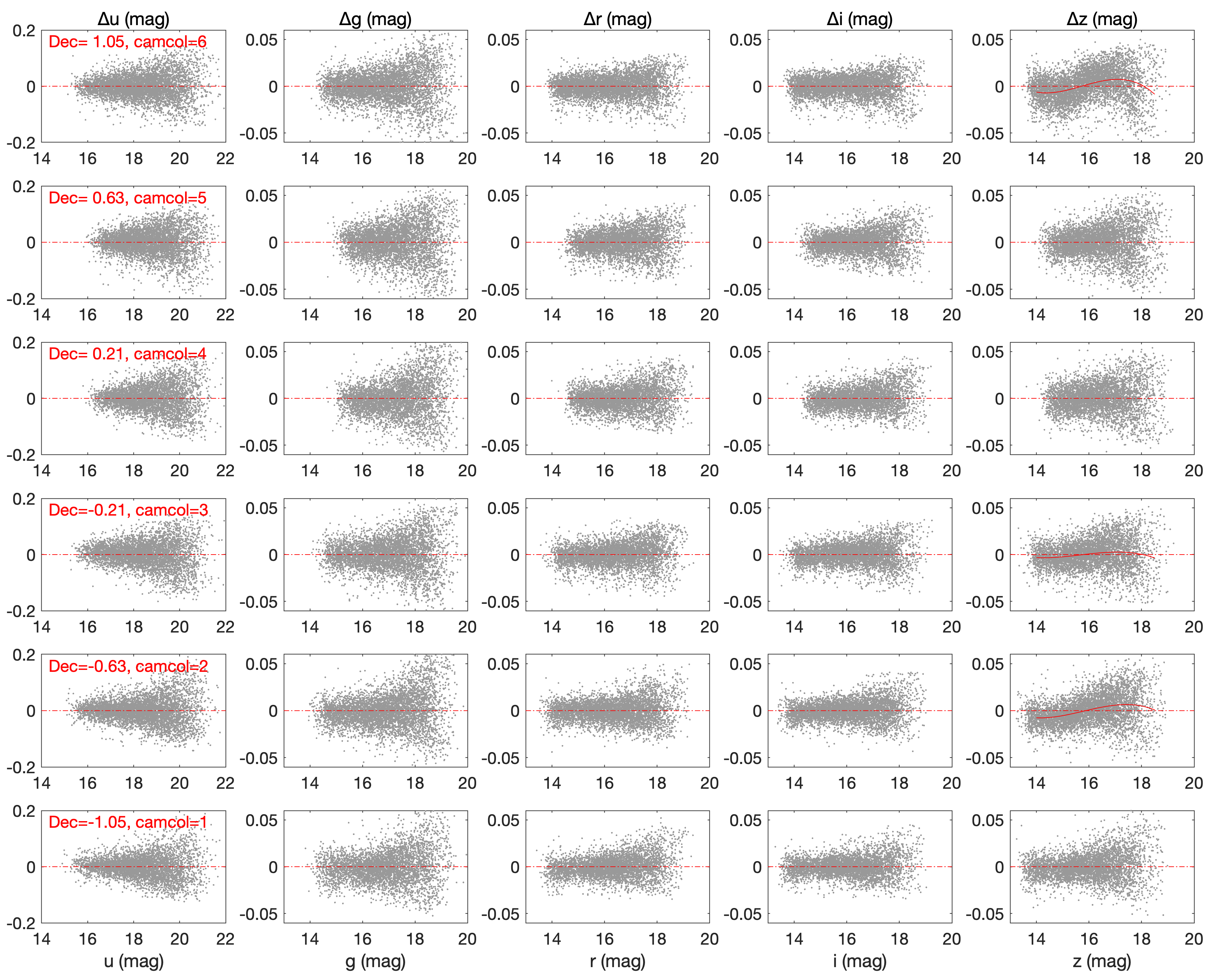}
\caption{The variations of magnitude offset as a function of magnitude for the six SDSS CCD columns after corrections of $\delta_m^{ext}({\rm R.A.})$ and $\delta_c^{ff}({\rm Dec.})$. The red solid lines are 3rd-order polynomial fitting results.}
\label{}
\end{figure*}

\begin{figure*}
\includegraphics[width=180mm]{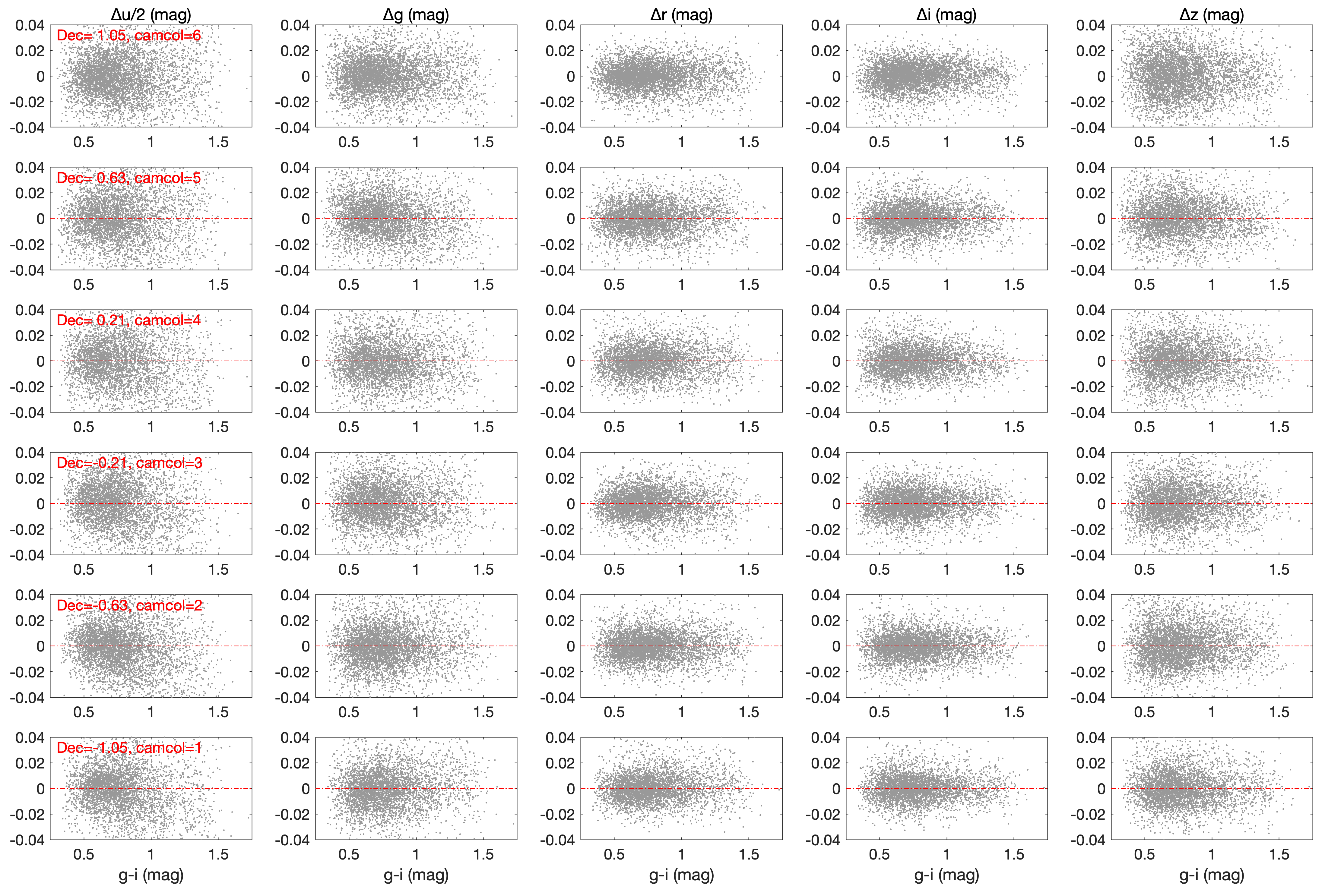}
\caption{The variations of magnitude offset as a function of color $g-i$ for the six SDSS CCD columns after corrections of $\delta_m^{ext}({\rm R.A.})$ and $\delta_c^{ff}({\rm Dec.})$.}
\label{}
\end{figure*}

\subsection{Final Precisions}

In this subsection, we apply three independent methods to estimate the precisions of our calibration results.

Firstly, the north and south strips were scanned  by the same CCDs. Flat field corrections are expected 
to be the same for the two strips. 
However, the $\delta_m^{ff}({\rm Dec.})$ of the two strips are inconsistent due to different flat field
corrections in I07. 
Taking into account the differences in I07, a good consistency between the two strips 
is found for each CCD. The small constant offsets between the two strips are probably caused by different zero points. The comparisons are shown in Figure\,16. The precisions of the $\delta_m^{ff}({\rm Dec})$ are estimated. 
The precision is about 7 mmag for the $u$ band. For the $g$, $r$, $i$, and $z$ bands, the precisions are about 2--3 mmag.

\begin{figure*}
\includegraphics[width=180mm]{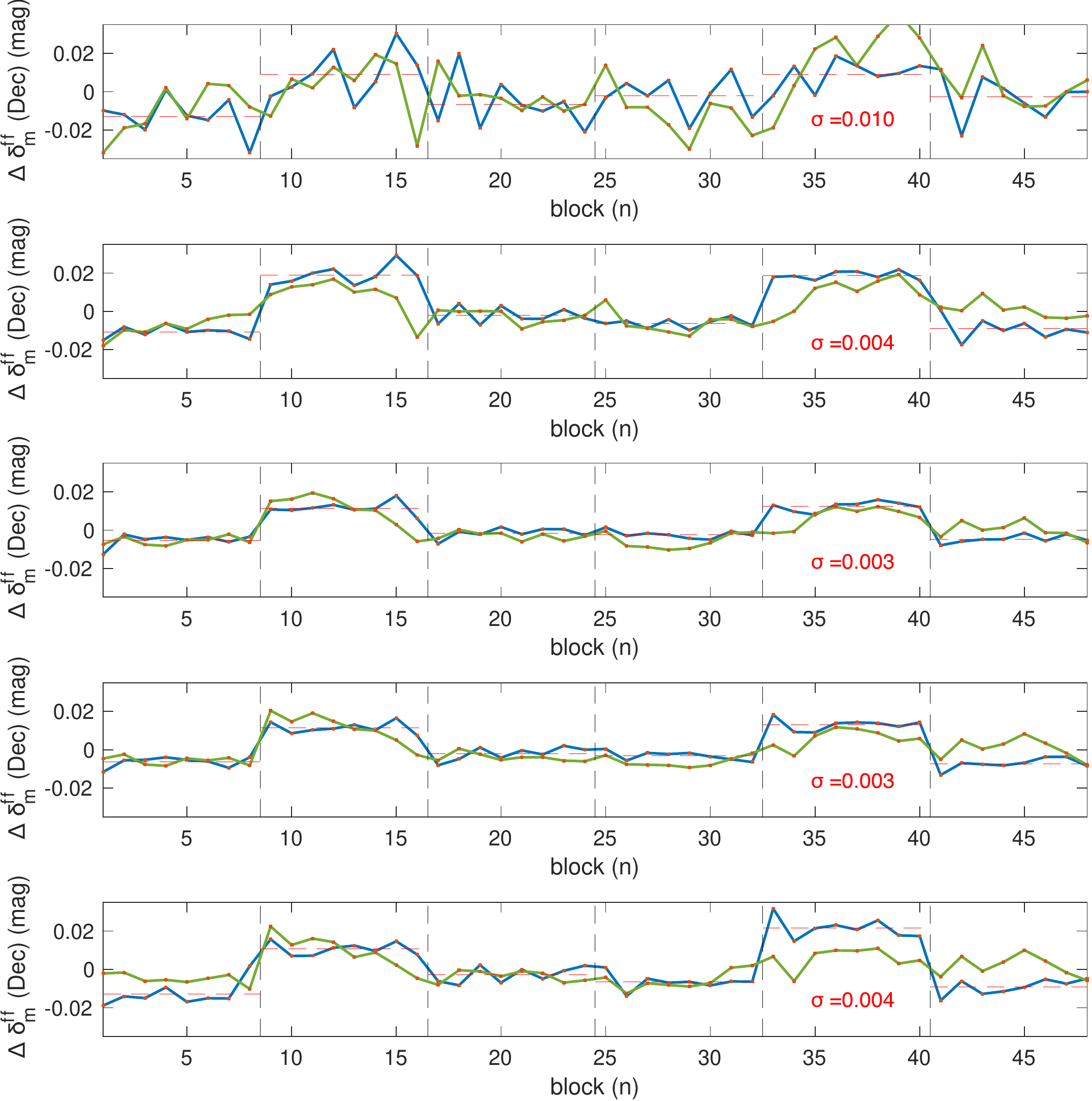}
\centering
\caption{The differences of $\delta_m^{ff}({\rm Dec.})$ between the two strips in the $u$, $g$, $r$, $i$, and $z$ bands (from top to bottom). The green solid lines mark the original differences. The blue solid lines mark the differences after accounting for the effect of flat fielding corrections of I07. The vertical dashed lines mark the approximate boundaries between different CCD columns. The red dashed line for each CCD column marks the 
mean value. The standard deviations of the blue lines with respect to the red lines are labeled.}
\label{}
\end{figure*}

Secondly, we compare the $\delta_m^{ext}({\rm R.A.})$ and $\delta_c^{ff}({\rm Dec.})$ results between the LAMOST and SDSS target samples.
Consistent results are found and shown in Figure\,17 for $\delta_m^{ext}({\rm R.A.})$  
and in Figure\,18 for $\delta_c^{ff}({\rm Dec.})$.
The comparisons suggest that the precisions are about 5--6 mmag in $u$ and 
about 2 mmag in the $g$, $r$, $i$, and $z$ bands.

\begin{figure*}
\includegraphics[width=180mm]{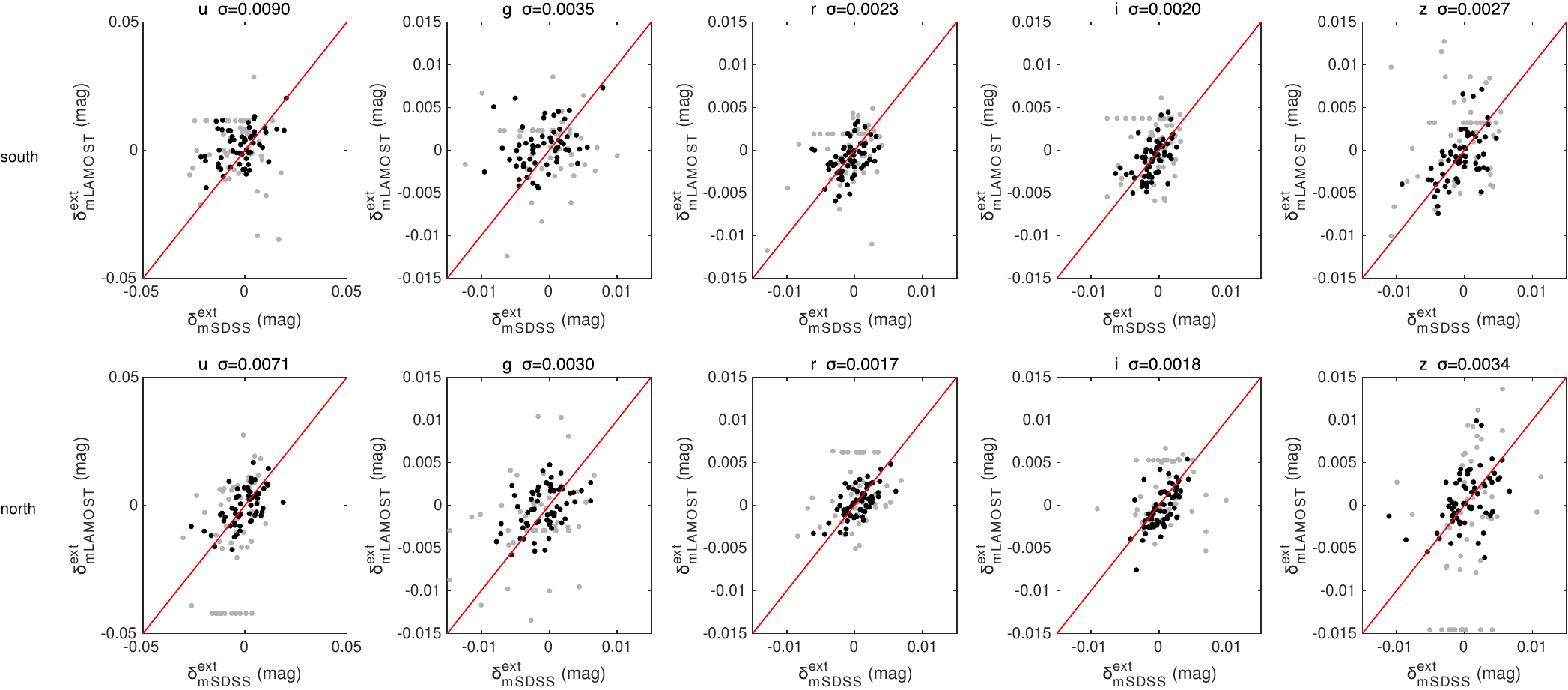}
\caption{The comparisons of $\delta_m^{ext}({\rm R.A.})$ for the south (top) and north (bottom) strips 
between the LAMOST and SDSS samples. The lines of equality are over-plotted in red.
Data points that are obtained with less than 80 stars are plotted in grey. 
The dispersion on the top of each panel considers only black points.}
\label{}
\end{figure*}

\begin{figure*}
\includegraphics[width=180mm]{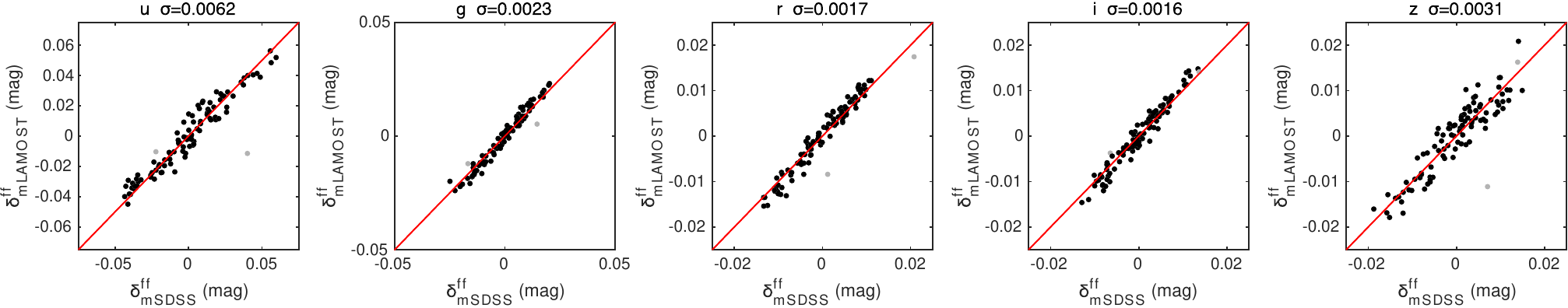}
\caption{The comparisons of $\delta_m^{ff}({\rm Dec.})$ between the LAMOST and SDSS samples. The lines of equality are over-plotted in red.
Data points that are obtained with less than 100 stars are plotted in grey. 
The dispersion on the top of each panel considers only black points.}
\label{}
\end{figure*}

Lastly, we plot the $\delta_c^{ff}({\rm Dec.})$ (derived from our $\delta_m^{ff}({\rm Dec.})$) versus $\delta_{g-r}^{ff}({\rm Dec.})$ in Figure\,19. Strong correlations are found between the color offsets,  and consistent with Yuan et al. (2015a). The slopes also agree well with the expected values (see 
more details in Section\,3.6 of Yuan et al. 2015a). The dispersions against the expected relations are consistent with the reported precisions. 

\begin{figure*}
\includegraphics[width=140mm]{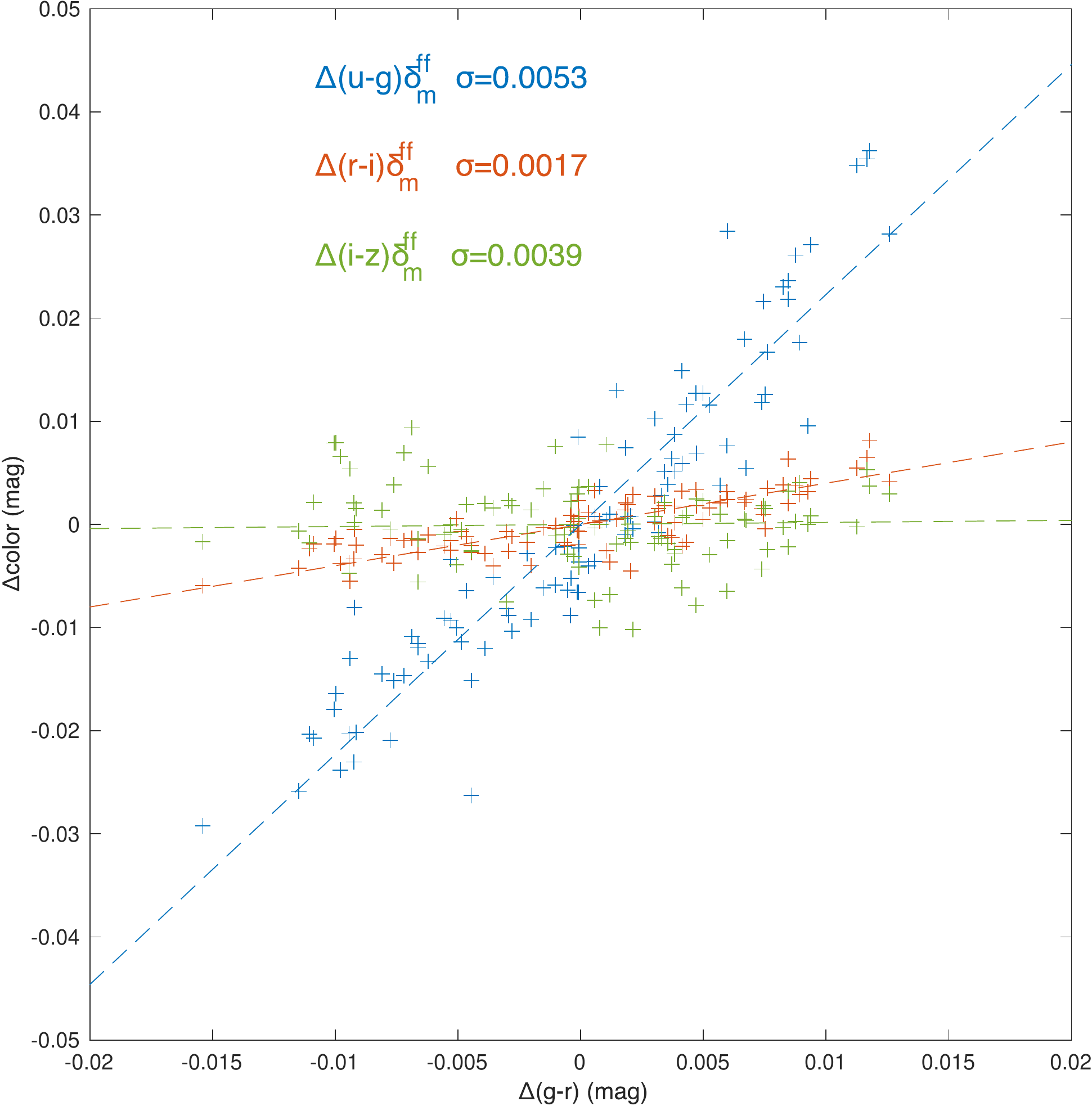}
\centering
\caption{$\delta_c^{ff}({\rm Dec.})$ versus $\delta_{g-r}^{ff}({\rm Dec.})$ for the LAMOST and SDSS samples. The dashed lines mark the expected relations for each color. The dispersions of $\delta_c^{ff}({\rm Dec.})$ against the expected relations are marked.}
\label{}
\end{figure*}

\section {Applying the method to the new version of SDSS Stripe\,82 standard stars catalog}
A new version (V4.2) of SDSS Stripe\,82 standard stars catalog is recently released (Thanjavur et al. 2021). 
Compared to the original version, the new version delivers averaged SDSS $ugriz$ photometry for nearly one million stars 
brighter than $\sim$22. Thanks to 2--3 times more measurements per star, their random errors are 1.4--1.7 times smaller 
than those in the original catalog. 
The new catalog is calibrated against {\it Gaia} EDR3, firstly using the $G$ photometry to derive grey photometric zeropoint corrections as functions of R.A. and Dec., then using the $BP-RP$ color to derive relative corrections in the $ugiz$ bands to the $r$ band. 
In this section, 
we apply our method to the new catalog with the same procedures. 
The $\delta_m^{ext}({\rm R.A.})$ and $\delta_c^{ff}({\rm Dec.})$ values obtained for the V4.2 catalog are displayed in Figure\,20 and listed in Tables\,A4--A6 in the Appendix. Same to Tables\,A1--A3, the entries in the tables should be added to cataloged photometry.

Two phenomena are found.
One is that the magnitude offsets in the Dec. direction are well corrected in the V4.2 catalog, particularly 
in the small scales within a given CCD camera.  
However, there are small (a few mmag) but significant global offsets for certain CCD cameras, e.g., 
the fourth camera in the south strip. 
The offsets for the fourth camera in the south strip are very similar in the $ugriz$ bands, probably due to the fact that the calibration of the $ugiz$ bands 
are based on the $r$ band and the offset comes from the $r$ band. 
The cause of such global offsets needs further investigation.

The other is that the magnitude offsets in the R.A. direction still suffer certain systematic, up to 0.01 -- 0.02 mag in the $ugriz$ bands. 
The errors are caused by ignoring the effects of varying reddening and metallicity along the R.A. direction in the calibration process of Thanjavur et al. (2021). Both reddening and metallicity increase 
toward low Galactic latitude regions.
The variations of reddening along the Dec. direction are very weak. Therefore, the calibration of Thanjavur et al. (2021) along the Dec. direction works well. 
For the $griz$ bands whose metallicity sensitivities are weak (about 0.02 mag/dex, see Yuan et al. 2015b), their errors are mainly 
caused by the effect of varying reddening. Consequently, their errors show a trend similar to extinction (see Figure\,1).
For the $u$ band whose metallicity sensitivity is strong (about 0.2 mag/dex, see Yuan et al. 2015b), its errors are caused by 
the combined effects of varying reddening and metallicity.

The variations of magnitude offset as a function of magnitude and color $g - i$ for the six SDSS CCD columns after corrections for V4.2 are similar to Figures\,14--15. 
Fit coefficients for the $z$ magnitude offsets as a function of $z$ magnitude are listed in Table\,3. The corrections should be added to the cataloged photometry.

\begin{table}
\centering
\caption{Fit coefficients for the $z$ magnitude offsets as a function of $z$ magnitude of the V4.2 catalog.}
\label{} \small
\begin{tabular}{rrrrr} \hline\hline
 Camcol & $z^3$ & $z^2$ & $z$ & Constant term\\ \hline
 2 & $-0.0006906$ & $0.03272$ & $-0.5108$ & $2.625$\\
 3 & $-0.0003347$ & $0.01599$ & $-0.2521$ & $1.310$\\
 6 & $-0.0003234$ & $0.01460$ & $-0.2161$ & $1.045$ \\
\hline
\end{tabular}
\end{table}

\begin{figure*}
\includegraphics[width=180mm]{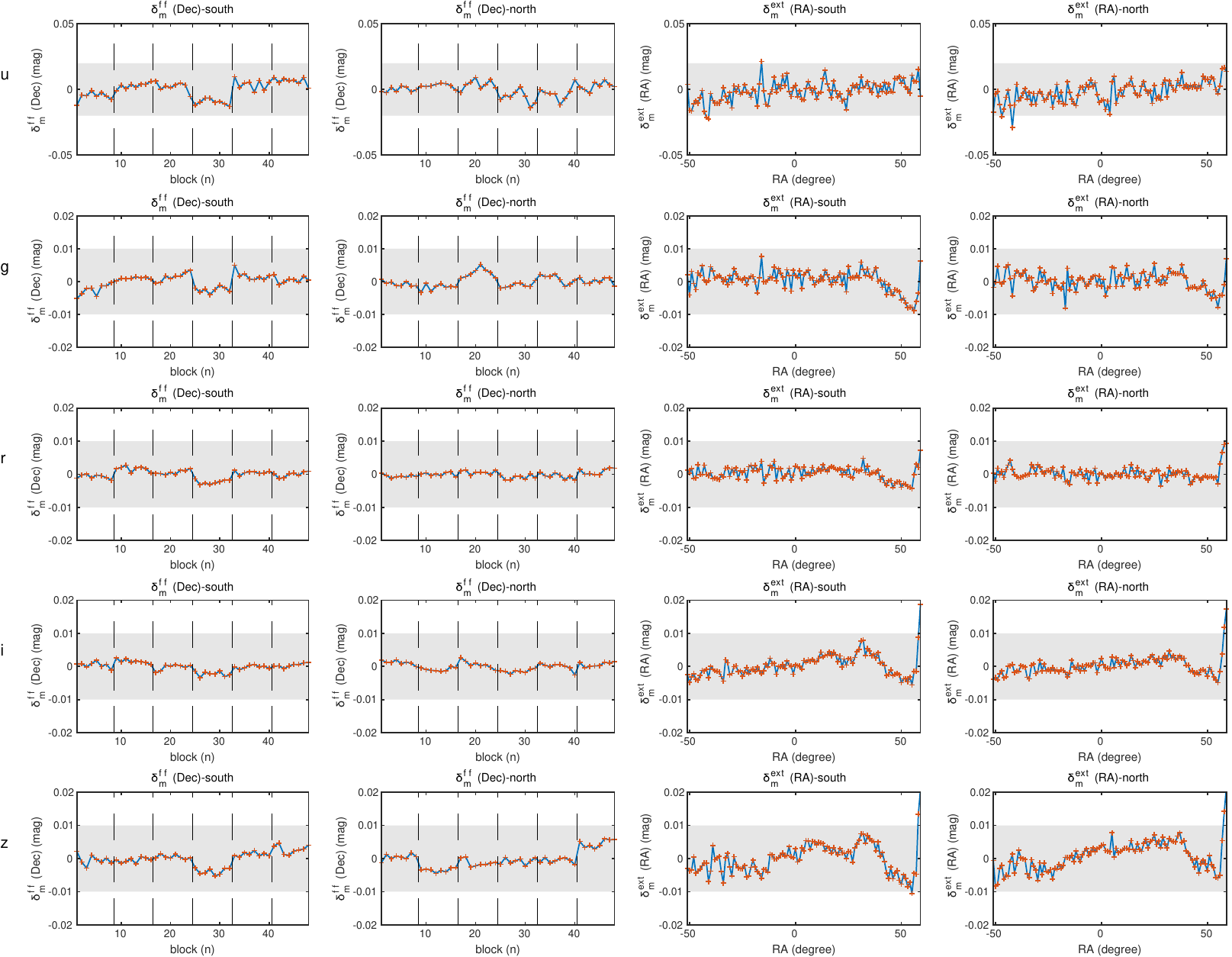}
\caption{Same to Figure\,9 but for the V4.2 catalog.}
\label{}
\end{figure*}

\section{Conclusions and discussions}
In this work, by combining spectroscopic data from the LAMOST DR7, SDSS DR12 and corrected photometric data from the {\it Gaia} EDR3, we apply the Stellar Color Regression method to recalibrate the SDSS Stripe 82 standard stars catalog. 
With a total number of about 30,000 spectroscopically targeted stars, we have mapped out the relatively large and strongly correlated photometric zero-point errors present in the catalog, $\sim$2.5 per cent in the $u$ band and $\sim$ 1 per cent in the $griz$ bands. Our study also confirms some small but significant magnitude dependence errors in the $z$ band for some CCDs. 
Various tests show that we have achieved an internal precision of about 5 mmag in the $u$ band and about 2 mmag in the $griz$ bands, which is about 5 times better than previous results. We also apply the method to the latest V4.2 version of the catalog, and find modest systematic calibration errors along the R.A. 
direction and smaller errors along the Dec. direction. The updated catalogs are publicly available\footnote{https://faculty.washington.edu/ivezic/sdss/catalogs/\\stripe82.html}\footnote{http://paperdata.china-vo.org/Huang.Bowen/2022/ApJS/\\PRoStripe82}\footnote{There are two catalogs in the link of the third footnote: the stripe82calibStars\_v4.2\_corrected.dat for V4.2 and the stripe82calibStars\_v2.6\_corrected.dat for I07.}.

The results demonstrate the power of the SCR method when combining spectroscopic data and \emph{Gaia} photometry in breaking the 1 percent precision barrier of ground-based photometric surveys. Our work paves the way for the re-calibration of existing surveys, such as the whole SDSS photometric survey, and has important implications for the calibration of future surveys, such as the LSST, CSST, and Mephisto.

The key idea behind the SCR method is that stellar colors are intrinsically simple, which can be fully determined by a small number of parameters including {\teff}, {\feh}, {\logg}, and other elemental abundances.
To make the method flexible under different situations, different forms can be adopted to predict intrinsic colors of the selected target stars:

 \begin{enumerate}
      \item  Intrinsic colors are functions of stellar atmospheric parameters. They can be computed via the star-pair technique 
or polynomial fitting relations.
Depending on the color of interest, different atmospheric parameters can be included. For broadband filters as in this work, 
{\teff}, {\feh} and {\logg} are sufficient. For some narrow-band 
filters such as $J510$ of the J-PLUS survey, one may need to include the effect of [Mg/Fe] as well. 

\item  Intrinsic colors are functions of normalized stellar spectra. One can use machine learning techniques to predict intrinsic 
colors of a given star from its normalized stellar spectrum. In this way, stellar atmospheric parameters are not needed anymore.  
This approach is pure data-driven and model-free.

\item  Intrinsic colors of main-sequence stars (or red giants if necessary) 
are functions of a given color (e.g., $BP-RP$) and metallicity via the tools of 
metallicity-dependent stellar locus (Yuan et al. 2015b; Huang et al. 2021a; 
L{\'o}pez-Sanjuan et al. 2021; Zhang et al. 2021). 
The metallicities of target stars could be from spectroscopic 
measurements when they are available,  or photometric ones delivered by well-calibrated photometric surveys (e.g., Yuan et al. 2015c; Huang et al. 2021b; Xu et al. 2021; Yang et al. 2022). 
It is worth mentioning that in the literature, stellar color-color relations 
are widely used as transformation relations between different photometric systems (e.g., Riello et al. 2021), 
ignoring the effects of metallicity and reddening. Although it is convenient, it will cause spatially dependent 
systematic errors up to a few per cent and should not be used when high-precision calibrations are required. 
However, using stars within a very limited sky area where variations of reddening and metallicity can be safely ignored, 
the color-color relations may serve as an excellent tool in correcting for small-scale effects (e.g., flat fielding).

\item  Intrinsic colors are functions of a given set of colors (e.g., $U-B$, $B-V$, $V-R$, and $R-I$) that are well-calibrated and
sensitive to stellar atmospheric parameters (e.g., Yang et al. 2021). It will also be very
interesting to explore predicting observed colors directly from the Gaia $BP$ and $RP$ spectrophotometry 
when the data is available. 
   \end{enumerate}

Alternatively,  with {\it Gaia} parallaxes accurate to a few per cent, 
and absolute magnitudes predicted from stellar spectra to 0.1 -- 0.2 magnitude 
(e.g., Xiang et al. 2017; Wang et al. submitted), it is also possible to 
predict the observed magnitudes directly based on {\it Gaia} parallaxes and spectra, without the help 
from {\it Gaia} photometry.  Such approach will be explored in future. 

Precise reddening correction is another key ingredient of the SCR method. Given that mmag precision has 
been achieved with the {\it Gaia} photometry, reddening correction precise to mmag also should be pursued with efforts. 
Three factors have to be considered:

\begin{enumerate}
      \item Systematics of widely used 2D reddening maps. Using millions of LAMOST stars, Sun et al. (submitted) have investigated 
      the SFD and \emph{Planck} 2D extinction maps (Planck collaboration 2014; Irfan et al. 2019) in the middle and high 
      Galactic latitude regions. Spatially dependent errors are found, which are correlated with the dust temperature, dust reddening, 
      and spectral index of the dust emission. Sun et al. (submitted) have provided recalibrated SFD and \emph{Planck} extinction maps 
      within the LAMOST footprint, along with empirical relations for regions outside. Nevertheless,  
      further improvements of the Galactic all-sky extinction maps are needed in the era of precision astronomy.

      \item Varying reddening coefficients for very broad or blue filters. For very broad (e.g., the {\it Gaia} passbands) or 
      blue (e.g., $u$, $NUV$, and $FUV$) filters, their reddening coefficients relative to \ebv~show strong dependences on stellar types and reddening (e.g., Niu et al. 2021a,b, Zhang et al. in preparation), 
      even for a given reddening curve. Such dependences should be carefully taken into account in future.

      \item Variations of reddening laws across the Galaxy, particularly in the Galactic disk. We will map the spatial variations of the reddening law across the Galactic disk in future (Zhang et al. in preparation).
\end{enumerate}    

Last but not least, the predicted magnitudes of the SCR method are for the ``standard" passbands
defined by the reference field. 
``Chromatic correction" (e.g., Burke et al. 2018) to the standard system is necessary in many cases
to account for the variations of passbands caused by atmospheric extinction and other factors.

\vspace{7mm} \noindent {\bf Acknowledgments}
{The authors thank the referee for his/her suggestions that improved the clarity of our presentation.
We acknowledge helpful discussions with Prof. Zeljko Ivezi{\'c} and Prof. Xiaowei Liu.
This work is supported by the National Natural Science Foundation of China through the project NSFC 12173007 and NSFC11603002,
the National Key Basic R\&D Program of China via 2019YFA0405503 and Beijing Normal University grant No. 310232102. 
We acknowledge the science research grants from the China Manned Space Project with NO. CMS-CSST-2021-A08 and CMS-CSST-2021-A09.

This work has made use of data from the European Space Agency (ESA) mission {\it Gaia} (\url{https://www.cosmos.esa.int/gaia}), processed by the Gaia Data Processing and Analysis Consortium (DPAC, \url{https:// www.cosmos.esa.int/web/gaia/dpac/ consortium}). Funding for the DPAC has been provided by national institutions, in particular the institutions participating in the Gaia Multilateral Agreement. 

Guoshoujing Telescope (the Large Sky Area Multi-Object Fiber Spectroscopic Telescope LAMOST) is a National Major Scientific Project built by the Chinese Academy of Sciences. Funding for the project has been provided by the National Development and Reform Commission. LAMOST is operated and managed by the National Astronomical Observatories, Chinese Academy of Sciences. 

Funding for SDSS-III has been provided by the Alfred P. Sloan Foundation, the Participating Institutions, the National Science Foundation, and the U.S.
Department of Energy Office of Science. The SDSS-III web site is http://www.sdss3.org/.SDSS-III is managed by the Astrophysical Research Consortium for the Participating Institutions of the SDSS-III Collaboration including the
University of Arizona, the Brazilian Participation Group, Brookhaven
NationalLaboratory, Carnegie Mellon University, University of Florida, the
French Participation Group, the German Participation Group, Harvard University, the
Instituto de Astrofisica de Canarias, the Michigan State/Notre Dame/JINA
Participation Group, Johns Hopkins University, Lawrence Berkeley National
Laboratory, Max Planck Institute for Astrophysics, Max Planck Institute for
Extraterrestrial Physics, New Mexico State University, New York University,
Ohio State University, Pennsylvania State University, University of Portsmouth,
Princeton University, the Spanish Participation Group, University of Tokyo,
University of Utah, Vanderbilt University, University of Virginia, University
of Washington, and Yale University.
}

{}

\clearpage

\appendix
\setcounter{table}{0}   
\setcounter{figure}{0}
\renewcommand{\thetable}{A\arabic{table}}
\renewcommand{\thefigure}{A\arabic{figure}}
\section {Magnitude offsets as a function of R.A. and Dec.}
\begin{table*}
\centering
\caption{$\delta_m^{ext}({\rm RA})$ of the north strip.}
\label{} \tiny 
\begin{tabular}{rrrrrrrrrrrrrrrrrr} \hline\hline
 RA   & $u$ &  $g$ & $r$  & $i$ & $z$ & RA & $u$ &  $g$ & $r$  & $i$ & $z$  & RA   & $u$ &  $g$ & $r$  & $i$ & $z$   \\
 (deg) & (mag) &  (mag) & (mag)  & (mag) & (mag) &(deg)   & (mag) &  (mag) & (mag)  & (mag) & (mag) & (deg)   & (mag) & (mag)& (mag) & (mag) & (mag)   \\\hline
 $-$50.5 & $-$0.015 & $-$0.006 & $+$0.001 & $+$0.001 & $-$0.002 & $-$13.5 & $-$0.004 & $+$0.003 & $+$0.002 & $+$0.001 & $+$0.002 & $+$23.5 & $-$0.000 & $-$0.003 & $-$0.004 & $-$0.001 & $-$0.005 \\
 $-$49.5 & $-$0.011 & $-$0.003 & $-$0.001 & $-$0.001 & $-$0.003 & $-$12.5 & $-$0.009 & $-$0.001 & $+$0.001 & $+$0.002 & $+$0.001 & $+$24.5 & $-$0.013 & $-$0.003 & $-$0.002 & $-$0.000 & $-$0.002 \\
 $-$48.5 & $-$0.007 & $+$0.003 & $+$0.003 & $-$0.002 & $-$0.005 & $-$11.5 & $-$0.009 & $-$0.005 & $+$0.001 & $-$0.000 & $+$0.003 & $+$25.5 & $-$0.000 & $+$0.001 & $-$0.002 & $-$0.003 & $-$0.000 \\
 $-$47.5 & $-$0.014 & $-$0.002 & $+$0.000 & $+$0.001 & $-$0.002 & $-$10.5 & $-$0.007 & $+$0.002 & $+$0.003 & $+$0.002 & $+$0.001 & $+$26.5 & $-$0.013 & $-$0.007 & $-$0.001 & $-$0.002 & $-$0.003 \\
 $-$46.5 & $-$0.017 & $-$0.001 & $+$0.001 & $+$0.003 & $-$0.001 & $-$9.5 & $+$0.007 & $+$0.002 & $+$0.002 & $+$0.002 & $+$0.003 & $+$27.5 & $-$0.002 & $-$0.001 & $-$0.001 & $-$0.001 & $-$0.001 \\
 $-$45.5 & $-$0.025 & $-$0.002 & $-$0.000 & $+$0.001 & $-$0.002 & $-$8.5 & $+$0.001 & $+$0.002 & $+$0.001 & $+$0.003 & $+$0.002 & $+$28.5 & $-$0.009 & $-$0.004 & $-$0.003 & $-$0.002 & $-$0.004 \\
 $-$44.5 & $-$0.006 & $+$0.006 & $+$0.003 & $-$0.002 & $-$0.005 & $-$7.5 & $+$0.000 & $-$0.000 & $-$0.001 & $+$0.001 & $+$0.004 & $+$29.5 & $+$0.003 & $-$0.000 & $-$0.002 & $-$0.003 & $-$0.002 \\
 $-$43.5 & $-$0.005 & $+$0.001 & $+$0.002 & $+$0.000 & $-$0.006 & $-$6.5 & $+$0.002 & $-$0.003 & $+$0.001 & $+$0.002 & $+$0.004 & $+$30.5 & $+$0.006 & $-$0.004 & $-$0.002 & $-$0.000 & $-$0.001 \\
 $-$42.5 & $-$0.015 & $-$0.006 & $+$0.000 & $+$0.000 & $+$0.001 & $-$5.5 & $+$0.002 & $-$0.002 & $-$0.001 & $+$0.001 & $+$0.003 & $+$31.5 & $-$0.009 & $-$0.005 & $-$0.004 & $-$0.005 & $-$0.007 \\
 $-$41.5 & $-$0.029 & $-$0.010 & $+$0.001 & $-$0.001 & $-$0.002 & $-$4.5 & $+$0.005 & $+$0.002 & $+$0.001 & $+$0.002 & $+$0.005 & $+$32.5 & $+$0.003 & $-$0.003 & $-$0.002 & $-$0.001 & $-$0.002 \\
 $-$40.5 & $-$0.013 & $-$0.001 & $-$0.001 & $-$0.001 & $-$0.004 & $-$3.5 & $+$0.002 & $+$0.002 & $+$0.003 & $+$0.003 & $+$0.004 & $+$33.5 & $+$0.002 & $-$0.002 & $-$0.002 & $-$0.002 & $-$0.002 \\
 $-$39.5 & $+$0.002 & $+$0.001 & $-$0.002 & $-$0.002 & $+$0.000 & $-$2.5 & $+$0.006 & $+$0.003 & $+$0.001 & $+$0.003 & $+$0.002 & $+$34.5 & $+$0.002 & $-$0.000 & $-$0.001 & $+$0.000 & $-$0.001 \\
 $-$38.5 & $-$0.008 & $-$0.000 & $-$0.002 & $-$0.001 & $+$0.004 & $-$1.5 & $-$0.015 & $-$0.001 & $-$0.000 & $+$0.002 & $+$0.006 & $+$35.5 & $+$0.002 & $+$0.001 & $+$0.000 & $-$0.001 & $+$0.000 \\
 $-$37.5 & $-$0.008 & $+$0.000 & $-$0.000 & $+$0.002 & $-$0.002 & $-$0.5 & $-$0.009 & $-$0.001 & $+$0.002 & $+$0.003 & $+$0.007 & $+$36.5 & $+$0.007 & $+$0.001 & $-$0.000 & $-$0.002 & $-$0.002 \\
 $-$36.5 & $-$0.002 & $-$0.000 & $-$0.000 & $+$0.001 & $-$0.001 & $+$0.5 & $-$0.007 & $-$0.003 & $-$0.001 & $+$0.003 & $+$0.004 & $+$37.5 & $+$0.004 & $+$0.002 & $-$0.000 & $-$0.000 & $+$0.001 \\
 $-$35.5 & $+$0.003 & $-$0.000 & $+$0.000 & $-$0.000 & $-$0.000 & $+$1.5 & $-$0.008 & $-$0.002 & $+$0.001 & $+$0.002 & $+$0.003 & $+$38.5 & $+$0.011 & $+$0.005 & $+$0.001 & $-$0.002 & $-$0.001 \\
 $-$34.5 & $-$0.007 & $-$0.003 & $-$0.001 & $-$0.002 & $-$0.003 & $+$2.5 & $-$0.009 & $-$0.005 & $+$0.000 & $+$0.003 & $+$0.004 & $+$39.5 & $+$0.002 & $+$0.001 & $+$0.000 & $-$0.001 & $+$0.001 \\
 $-$33.5 & $-$0.003 & $+$0.000 & $-$0.001 & $+$0.000 & $-$0.000 & $+$3.5 & $-$0.010 & $-$0.002 & $-$0.000 & $+$0.002 & $+$0.003 & $+$40.5 & $+$0.004 & $+$0.002 & $+$0.000 & $-$0.002 & $-$0.003 \\
 $-$32.5 & $-$0.014 & $-$0.002 & $+$0.001 & $+$0.001 & $-$0.002 & $+$4.5 & $-$0.017 & $-$0.000 & $+$0.001 & $+$0.002 & $+$0.003 & $+$41.5 & $+$0.003 & $-$0.001 & $+$0.000 & $-$0.002 & $-$0.000 \\
 $-$31.5 & $-$0.007 & $-$0.004 & $+$0.001 & $+$0.001 & $-$0.004 & $+$5.5 & $+$0.001 & $+$0.004 & $+$0.001 & $+$0.004 & $+$0.006 & $+$42.5 & $+$0.004 & $-$0.000 & $-$0.001 & $-$0.002 & $+$0.000 \\
 $-$30.5 & $-$0.013 & $-$0.002 & $+$0.001 & $+$0.002 & $-$0.001 & $+$6.5 & $+$0.002 & $+$0.003 & $+$0.001 & $+$0.000 & $-$0.001 & $+$43.5 & $+$0.002 & $-$0.002 & $+$0.000 & $-$0.001 & $+$0.004 \\
 $-$29.5 & $-$0.007 & $-$0.000 & $+$0.003 & $+$0.001 & $-$0.002 & $+$7.5 & $-$0.015 & $-$0.001 & $+$0.003 & $+$0.005 & $+$0.004 & $+$44.5 & $+$0.004 & $+$0.000 & $+$0.002 & $-$0.000 & $+$0.000 \\
 $-$28.5 & $-$0.001 & $+$0.004 & $+$0.005 & $+$0.004 & $+$0.002 & $+$8.5 & $-$0.005 & $+$0.003 & $+$0.001 & $+$0.001 & $-$0.000 & $+$45.5 & $+$0.004 & $-$0.002 & $+$0.001 & $-$0.002 & $+$0.001 \\
 $-$27.5 & $-$0.006 & $-$0.000 & $+$0.002 & $+$0.001 & $-$0.002 & $+$9.5 & $-$0.004 & $+$0.001 & $+$0.000 & $+$0.001 & $+$0.000 & $+$46.5 & $+$0.005 & $+$0.001 & $+$0.001 & $-$0.001 & $-$0.001 \\
 $-$26.5 & $-$0.007 & $+$0.002 & $+$0.003 & $+$0.005 & $+$0.002 & $+$10.5 & $-$0.005 & $-$0.003 & $-$0.001 & $+$0.001 & $-$0.000 & $+$47.5 & $+$0.003 & $-$0.001 & $-$0.001 & $-$0.001 & $-$0.001 \\
 $-$25.5 & $-$0.006 & $-$0.000 & $+$0.002 & $+$0.002 & $+$0.000 & $+$11.5 & $+$0.001 & $-$0.000 & $+$0.001 & $+$0.002 & $+$0.001 & $+$48.5 & $+$0.009 & $+$0.001 & $+$0.001 & $-$0.000 & $+$0.000 \\
 $-$24.5 & $-$0.006 & $+$0.000 & $+$0.002 & $+$0.002 & $+$0.000 & $+$12.5 & $+$0.006 & $+$0.005 & $+$0.000 & $-$0.001 & $+$0.001 & $+$49.5 & $+$0.013 & $+$0.002 & $+$0.001 & $-$0.000 & $+$0.002 \\
 $-$23.5 & $-$0.003 & $+$0.000 & $+$0.002 & $+$0.001 & $+$0.000 & $+$13.5 & $+$0.005 & $+$0.001 & $-$0.002 & $+$0.001 & $+$0.000 & $+$50.5 & $+$0.009 & $+$0.000 & $+$0.001 & $-$0.001 & $+$0.002 \\
 $-$22.5 & $-$0.014 & $-$0.003 & $+$0.004 & $+$0.004 & $-$0.001 & $+$14.5 & $+$0.007 & $+$0.001 & $+$0.003 & $+$0.002 & $+$0.003 & $+$51.5 & $+$0.006 & $-$0.000 & $+$0.001 & $+$0.002 & $+$0.005 \\
 $-$21.5 & $-$0.006 & $-$0.002 & $-$0.000 & $+$0.000 & $-$0.001 & $+$15.5 & $+$0.001 & $+$0.002 & $+$0.002 & $+$0.002 & $+$0.003 & $+$52.5 & $+$0.006 & $-$0.001 & $+$0.001 & $+$0.001 & $+$0.003 \\
 $-$20.5 & $+$0.001 & $-$0.001 & $+$0.002 & $+$0.002 & $+$0.001 & $+$16.5 & $-$0.001 & $-$0.005 & $+$0.000 & $-$0.000 & $+$0.000 & $+$53.5 & $+$0.009 & $+$0.002 & $+$0.002 & $+$0.000 & $+$0.003 \\
 $-$19.5 & $-$0.008 & $-$0.003 & $+$0.001 & $+$0.001 & $+$0.001 & $+$17.5 & $-$0.006 & $-$0.001 & $-$0.000 & $-$0.001 & $+$0.000 & $+$54.5 & $+$0.002 & $-$0.001 & $+$0.003 & $-$0.001 & $-$0.001 \\
 $-$18.5 & $-$0.004 & $-$0.004 & $-$0.000 & $+$0.002 & $+$0.002 & $+$18.5 & $-$0.006 & $-$0.002 & $-$0.001 & $-$0.001 & $+$0.000 & $+$55.5 & $+$0.002 & $-$0.002 & $+$0.000 & $-$0.001 & $-$0.001 \\
 $-$17.5 & $-$0.012 & $-$0.004 & $+$0.001 & $+$0.003 & $+$0.000 & $+$19.5 & $-$0.000 & $+$0.002 & $+$0.001 & $+$0.000 & $+$0.002 & $+$56.5 & $-$0.002 & $+$0.002 & $+$0.004 & $+$0.000 & $-$0.001 \\
 $-$16.5 & $-$0.014 & $-$0.009 & $-$0.002 & $+$0.003 & $+$0.004 & $+$20.5 & $+$0.001 & $+$0.001 & $+$0.000 & $-$0.000 & $+$0.001 & $+$57.5 & $+$0.004 & $-$0.004 & $-$0.000 & $-$0.001 & $+$0.005 \\
 $-$15.5 & $+$0.004 & $+$0.004 & $+$0.001 & $+$0.002 & $+$0.004 & $+$21.5 & $-$0.000 & $-$0.005 & $-$0.003 & $-$0.002 & $-$0.000 & $+$58.5 & $-$0.001 & $-$0.010 & $-$0.004 & $-$0.002 & $-$0.000 \\
 $-$14.5 & $-$0.001 & $-$0.003 & $-$0.001 & $-$0.000 & $+$0.001 & $+$22.5 & $-$0.003 & $-$0.006 & $-$0.005 & $-$0.004 & $-$0.004 & $+$59.5 & $-$0.005 & $-$0.008 & $-$0.008 & $-$0.003 & $-$0.007 \\

\hline
\end{tabular}
\end{table*}

\begin{table*}
\centering
\caption{$\delta_m^{ext}({\rm RA})$ of the south strip.}
\label{} \tiny 
\begin{tabular}{rrrrrrrrrrrrrrrrrr} \hline\hline
 RA   & $u$ &  $g$ & $r$  & $i$ & $z$ & RA & $u$ &  $g$ & $r$  & $i$ & $z$  & RA   & $u$ &  $g$ & $r$  & $i$ & $z$   \\
 (deg) & (mag) &  (mag) & (mag)  & (mag) & (mag) &(deg)   & (mag) &  (mag) & (mag)  & (mag) & (mag) & (deg)   & (mag) & (mag)& (mag) & (mag) & (mag)   \\\hline
 $-$50.5 & $+$0.001 & $-$0.001 & $-$0.004 & $+$0.002 & $-$0.000 & $-$13.5 & $-$0.008 & $-$0.002 & $-$0.001 & $+$0.001 & $-$0.004 & $+$23.5 & $-$0.011 & $-$0.004 & $-$0.003 & $-$0.005 & $-$0.007 \\
 $-$49.5 & $-$0.024 & $-$0.006 & $+$0.000 & $-$0.004 & $+$0.001 & $-$12.5 & $-$0.007 & $+$0.001 & $-$0.001 & $+$0.001 & $-$0.004 & $+$24.5 & $-$0.018 & $-$0.007 & $-$0.006 & $-$0.006 & $-$0.005 \\
 $-$48.5 & $-$0.018 & $-$0.002 & $+$0.001 & $+$0.000 & $+$0.002 & $-$11.5 & $+$0.000 & $+$0.002 & $+$0.002 & $+$0.003 & $+$0.001 & $+$25.5 & $-$0.004 & $-$0.004 & $-$0.002 & $-$0.001 & $-$0.005 \\
 $-$47.5 & $-$0.014 & $-$0.003 & $-$0.005 & $-$0.002 & $+$0.001 & $-$10.5 & $+$0.003 & $+$0.004 & $+$0.003 & $+$0.003 & $+$0.002 & $+$26.5 & $-$0.006 & $-$0.002 & $-$0.003 & $-$0.001 & $-$0.005 \\
 $-$46.5 & $-$0.007 & $-$0.004 & $-$0.004 & $-$0.005 & $+$0.002 & $-$9.5 & $+$0.005 & $+$0.006 & $+$0.003 & $+$0.003 & $+$0.002 & $+$27.5 & $+$0.002 & $-$0.003 & $-$0.003 & $-$0.005 & $-$0.010 \\
 $-$45.5 & $-$0.008 & $+$0.002 & $-$0.001 & $-$0.000 & $+$0.003 & $-$8.5 & $-$0.002 & $-$0.001 & $-$0.003 & $-$0.002 & $+$0.001 & $+$28.5 & $-$0.002 & $-$0.000 & $-$0.001 & $-$0.002 & $-$0.005 \\
 $-$44.5 & $-$0.005 & $-$0.000 & $-$0.001 & $-$0.003 & $-$0.003 & $-$7.5 & $+$0.006 & $+$0.003 & $+$0.001 & $+$0.001 & $+$0.002 & $+$29.5 & $+$0.001 & $+$0.001 & $-$0.002 & $-$0.002 & $-$0.003 \\
 $-$43.5 & $-$0.008 & $+$0.001 & $-$0.002 & $-$0.002 & $+$0.001 & $-$6.5 & $+$0.003 & $+$0.001 & $+$0.001 & $-$0.000 & $+$0.001 & $+$30.5 & $-$0.004 & $-$0.003 & $-$0.002 & $+$0.000 & $-$0.002 \\
 $-$42.5 & $-$0.018 & $-$0.002 & $+$0.001 & $-$0.000 & $-$0.002 & $-$5.5 & $+$0.009 & $+$0.002 & $-$0.001 & $+$0.002 & $+$0.003 & $+$31.5 & $+$0.008 & $+$0.002 & $-$0.002 & $+$0.000 & $-$0.000 \\
 $-$41.5 & $-$0.023 & $-$0.006 & $-$0.003 & $+$0.002 & $-$0.001 & $-$4.5 & $+$0.004 & $-$0.001 & $-$0.001 & $+$0.001 & $+$0.002 & $+$32.5 & $-$0.002 & $+$0.001 & $+$0.001 & $+$0.002 & $-$0.000 \\
 $-$40.5 & $-$0.022 & $-$0.004 & $-$0.003 & $-$0.004 & $-$0.005 & $-$3.5 & $+$0.006 & $+$0.003 & $-$0.001 & $+$0.001 & $+$0.001 & $+$33.5 & $+$0.003 & $-$0.001 & $-$0.002 & $-$0.001 & $-$0.001 \\
 $-$39.5 & $-$0.003 & $+$0.001 & $-$0.002 & $-$0.001 & $-$0.004 & $-$2.5 & $-$0.006 & $-$0.001 & $+$0.001 & $+$0.002 & $+$0.004 & $+$34.5 & $+$0.005 & $+$0.001 & $-$0.000 & $-$0.000 & $-$0.001 \\
 $-$38.5 & $-$0.010 & $+$0.000 & $-$0.004 & $-$0.001 & $+$0.002 & $-$1.5 & $-$0.007 & $-$0.003 & $-$0.002 & $+$0.003 & $+$0.003 & $+$35.5 & $+$0.001 & $-$0.002 & $-$0.002 & $-$0.001 & $-$0.000 \\
 $-$37.5 & $-$0.012 & $-$0.001 & $-$0.003 & $-$0.002 & $+$0.001 & $-$0.5 & $-$0.001 & $+$0.003 & $-$0.001 & $-$0.001 & $-$0.001 & $+$36.5 & $-$0.002 & $-$0.003 & $-$0.002 & $-$0.002 & $-$0.001 \\
 $-$36.5 & $-$0.011 & $-$0.000 & $-$0.003 & $-$0.003 & $+$0.001 & $+$0.5 & $+$0.006 & $+$0.000 & $+$0.001 & $+$0.002 & $+$0.005 & $+$37.5 & $+$0.001 & $+$0.002 & $-$0.001 & $-$0.001 & $-$0.001 \\
 $-$35.5 & $-$0.009 & $+$0.001 & $-$0.003 & $-$0.004 & $+$0.000 & $+$1.5 & $-$0.012 & $-$0.003 & $+$0.001 & $-$0.001 & $-$0.000 & $+$38.5 & $+$0.005 & $+$0.003 & $-$0.001 & $-$0.001 & $-$0.005 \\
 $-$34.5 & $-$0.009 & $-$0.001 & $-$0.002 & $-$0.003 & $-$0.002 & $+$2.5 & $+$0.002 & $+$0.002 & $+$0.002 & $+$0.001 & $-$0.001 & $+$39.5 & $-$0.002 & $+$0.001 & $+$0.001 & $+$0.001 & $-$0.001 \\
 $-$33.5 & $+$0.001 & $-$0.001 & $-$0.004 & $-$0.003 & $-$0.004 & $+$3.5 & $-$0.007 & $-$0.003 & $-$0.001 & $+$0.001 & $+$0.002 & $+$40.5 & $+$0.005 & $+$0.002 & $-$0.004 & $-$0.003 & $-$0.003 \\
 $-$32.5 & $-$0.004 & $-$0.002 & $-$0.001 & $+$0.001 & $-$0.001 & $+$4.5 & $-$0.008 & $+$0.001 & $+$0.002 & $+$0.002 & $+$0.004 & $+$41.5 & $+$0.001 & $-$0.001 & $-$0.003 & $-$0.004 & $-$0.001 \\
 $-$31.5 & $-$0.006 & $-$0.003 & $-$0.001 & $+$0.000 & $+$0.001 & $+$5.5 & $+$0.007 & $+$0.003 & $-$0.000 & $+$0.000 & $+$0.003 & $+$42.5 & $+$0.009 & $+$0.003 & $+$0.000 & $-$0.002 & $-$0.002 \\
 $-$30.5 & $-$0.008 & $+$0.000 & $-$0.001 & $+$0.000 & $+$0.000 & $+$6.5 & $+$0.011 & $+$0.003 & $+$0.001 & $+$0.001 & $+$0.004 & $+$43.5 & $+$0.000 & $-$0.002 & $-$0.001 & $-$0.001 & $-$0.002 \\
 $-$29.5 & $-$0.010 & $+$0.002 & $+$0.002 & $-$0.000 & $+$0.001 & $+$7.5 & $+$0.002 & $+$0.000 & $-$0.001 & $+$0.001 & $+$0.000 & $+$44.5 & $+$0.005 & $+$0.001 & $-$0.001 & $-$0.001 & $-$0.001 \\
 $-$28.5 & $+$0.002 & $+$0.002 & $-$0.001 & $-$0.000 & $-$0.001 & $+$8.5 & $-$0.002 & $+$0.001 & $+$0.001 & $+$0.002 & $+$0.004 & $+$45.5 & $-$0.002 & $-$0.004 & $-$0.001 & $-$0.001 & $-$0.000 \\
 $-$27.5 & $+$0.000 & $+$0.003 & $+$0.002 & $+$0.001 & $-$0.001 & $+$9.5 & $+$0.002 & $-$0.001 & $-$0.001 & $+$0.001 & $+$0.003 & $+$46.5 & $-$0.000 & $-$0.003 & $-$0.002 & $-$0.001 & $-$0.003 \\
 $-$26.5 & $+$0.001 & $+$0.002 & $-$0.000 & $+$0.001 & $+$0.001 & $+$10.5 & $-$0.008 & $-$0.001 & $-$0.002 & $+$0.001 & $+$0.002 & $+$47.5 & $+$0.002 & $-$0.001 & $-$0.003 & $-$0.003 & $-$0.004 \\
 $-$25.5 & $-$0.003 & $+$0.002 & $+$0.001 & $+$0.001 & $+$0.002 & $+$11.5 & $-$0.001 & $-$0.001 & $-$0.002 & $+$0.000 & $+$0.002 & $+$48.5 & $+$0.009 & $+$0.000 & $-$0.001 & $-$0.002 & $-$0.003 \\
 $-$24.5 & $-$0.003 & $+$0.001 & $+$0.001 & $-$0.000 & $+$0.002 & $+$12.5 & $-$0.001 & $+$0.003 & $+$0.000 & $+$0.002 & $+$0.003 & $+$49.5 & $+$0.014 & $-$0.000 & $-$0.002 & $-$0.002 & $-$0.004 \\
 $-$23.5 & $+$0.003 & $+$0.002 & $+$0.001 & $+$0.001 & $+$0.003 & $+$13.5 & $+$0.006 & $+$0.001 & $+$0.002 & $+$0.002 & $+$0.003 & $+$50.5 & $+$0.006 & $-$0.002 & $-$0.002 & $-$0.003 & $-$0.003 \\
 $-$22.5 & $-$0.003 & $+$0.001 & $-$0.001 & $+$0.001 & $+$0.000 & $+$14.5 & $+$0.014 & $+$0.002 & $+$0.001 & $+$0.000 & $+$0.000 & $+$51.5 & $+$0.003 & $-$0.002 & $-$0.003 & $-$0.003 & $-$0.002 \\
 $-$21.5 & $-$0.003 & $-$0.000 & $+$0.000 & $-$0.001 & $-$0.001 & $+$15.5 & $+$0.001 & $+$0.002 & $+$0.001 & $+$0.002 & $+$0.001 & $+$52.5 & $+$0.002 & $-$0.003 & $-$0.002 & $-$0.002 & $-$0.001 \\
 $-$20.5 & $-$0.002 & $+$0.001 & $+$0.002 & $+$0.000 & $+$0.001 & $+$16.5 & $+$0.001 & $+$0.000 & $+$0.001 & $+$0.003 & $+$0.001 & $+$53.5 & $+$0.000 & $-$0.003 & $-$0.003 & $-$0.003 & $-$0.004 \\
 $-$19.5 & $+$0.003 & $+$0.001 & $+$0.000 & $-$0.001 & $+$0.003 & $+$17.5 & $-$0.009 & $-$0.002 & $-$0.000 & $-$0.001 & $-$0.003 & $+$54.5 & $-$0.003 & $-$0.005 & $-$0.001 & $-$0.002 & $-$0.005 \\
 $-$18.5 & $-$0.002 & $+$0.000 & $+$0.002 & $+$0.000 & $+$0.001 & $+$18.5 & $+$0.008 & $+$0.003 & $+$0.001 & $-$0.000 & $+$0.000 & $+$55.5 & $+$0.013 & $-$0.003 & $-$0.004 & $-$0.003 & $-$0.005 \\
 $-$17.5 & $-$0.016 & $-$0.006 & $+$0.000 & $+$0.001 & $-$0.001 & $+$19.5 & $+$0.001 & $-$0.001 & $+$0.001 & $+$0.000 & $-$0.001 & $+$56.5 & $-$0.000 & $-$0.003 & $-$0.003 & $-$0.003 & $-$0.001 \\
 $-$16.5 & $+$0.011 & $+$0.002 & $+$0.003 & $+$0.000 & $-$0.001 & $+$20.5 & $-$0.005 & $-$0.001 & $-$0.000 & $-$0.002 & $-$0.003 & $+$57.5 & $-$0.006 & $-$0.005 & $-$0.005 & $-$0.004 & $-$0.002 \\
 $-$15.5 & $+$0.020 & $+$0.008 & $+$0.002 & $-$0.001 & $-$0.001 & $+$21.5 & $-$0.007 & $-$0.001 & $-$0.000 & $-$0.002 & $-$0.003 & $+$58.5 & $+$0.003 & $-$0.011 & $-$0.012 & $-$0.004 & $-$0.002 \\
 $-$14.5 & $-$0.001 & $+$0.001 & $-$0.001 & $-$0.001 & $+$0.002 & $+$22.5 & $-$0.008 & $-$0.004 & $-$0.005 & $-$0.005 & $-$0.003 & $+$59.5 & $-$0.017 & $-$0.009 & $-$0.011 & $-$0.005 & $-$0.011 \\

\hline
\end{tabular}
\end{table*}

\begin{table*}
\centering
\caption{$\delta_m^{ff}({\rm Dec})$.}
\label{} \tiny 
\begin{tabular}{rrrrrrrrrrrrrrrrrr} \hline\hline
 Dec   & $u$ &  $g$ & $r$  & $i$ & $z$ & Dec & $u$ &  $g$ & $r$  & $i$ & $z$  & Dec   & $u$ &  $g$ & $r$  & $i$ & $z$   \\
 (deg) & (mag) &  (mag) & (mag)  & (mag) & (mag) &(deg)   & (mag) &  (mag) & (mag)  & (mag) & (mag) & (deg)   & (mag) & (mag)& (mag) & (mag) & (mag)
 \\\hline
 $-$1.262 & $-$0.008 & $-$0.011 & $-$0.007 & $-$0.004 & $-$0.009 & $-$0.402 & $+$0.014 & $+$0.007 & $+$0.003 & $+$0.001 & $-$0.002 & $+$0.454 & $+$0.040 & $+$0.013 & $+$0.003 & $-$0.001 & $-$0.002 \\
 $-$1.232 & $-$0.036 & $-$0.023 & $-$0.013 & $-$0.008 & $-$0.013 & $-$0.376 & $+$0.006 & $+$0.007 & $+$0.005 & $+$0.003 & $+$0.003 & $+$0.480 & $+$0.045 & $+$0.017 & $+$0.005 & $+$0.000 & $-$0.002 \\
 $-$1.206 & $-$0.033 & $-$0.018 & $-$0.011 & $-$0.009 & $-$0.016 & $-$0.350 & $+$0.013 & $+$0.009 & $+$0.004 & $+$0.001 & $+$0.001 & $+$0.506 & $+$0.056 & $+$0.021 & $+$0.009 & $+$0.003 & $-$0.002 \\
 $-$1.180 & $-$0.037 & $-$0.021 & $-$0.011 & $-$0.010 & $-$0.017 & $-$0.324 & $+$0.022 & $+$0.012 & $+$0.003 & $-$0.000 & $-$0.000 & $+$0.532 & $+$0.038 & $+$0.017 & $+$0.009 & $+$0.008 & $+$0.006 \\
 $-$1.154 & $-$0.036 & $-$0.021 & $-$0.013 & $-$0.010 & $-$0.014 & $-$0.298 & $+$0.008 & $+$0.003 & $-$0.001 & $-$0.001 & $+$0.001 & $+$0.558 & $+$0.044 & $+$0.018 & $+$0.010 & $+$0.006 & $+$0.005 \\
 $-$1.128 & $-$0.038 & $-$0.020 & $-$0.010 & $-$0.008 & $-$0.016 & $-$0.272 & $+$0.022 & $+$0.007 & $+$0.003 & $-$0.000 & $+$0.006 & $+$0.584 & $+$0.055 & $+$0.021 & $+$0.009 & $+$0.004 & $+$0.004 \\
 $-$1.102 & $-$0.031 & $-$0.016 & $-$0.008 & $-$0.005 & $-$0.007 & $-$0.246 & $+$0.015 & $+$0.007 & $+$0.001 & $-$0.002 & $+$0.004 & $+$0.610 & $+$0.042 & $+$0.014 & $+$0.008 & $+$0.005 & $+$0.007 \\
 $-$1.076 & $-$0.035 & $-$0.015 & $-$0.004 & $-$0.002 & $+$0.000 & $-$0.220 & $+$0.006 & $+$0.006 & $+$0.004 & $+$0.001 & $+$0.011 & $+$0.635 & $+$0.054 & $+$0.018 & $+$0.006 & $-$0.002 & $-$0.006 \\
 $-$1.050 & $-$0.042 & $-$0.015 & $-$0.011 & $-$0.009 & $-$0.006 & $-$0.195 & $-$0.002 & $+$0.007 & $+$0.007 & $+$0.006 & $+$0.006 & $+$0.661 & $+$0.037 & $+$0.013 & $+$0.004 & $+$0.002 & $+$0.004 \\
 $-$1.025 & $-$0.004 & $-$0.005 & $-$0.006 & $-$0.003 & $-$0.011 & $-$0.169 & $+$0.008 & $+$0.007 & $+$0.005 & $+$0.003 & $+$0.003 & $+$0.687 & $+$0.023 & $+$0.005 & $-$0.004 & $-$0.007 & $-$0.011 \\
 $-$0.999 & $-$0.014 & $-$0.008 & $-$0.007 & $-$0.007 & $-$0.015 & $-$0.143 & $+$0.015 & $+$0.009 & $+$0.006 & $+$0.004 & $+$0.002 & $+$0.713 & $+$0.028 & $+$0.006 & $-$0.003 & $-$0.009 & $-$0.012 \\
 $-$0.973 & $-$0.021 & $-$0.010 & $-$0.003 & $-$0.002 & $-$0.011 & $-$0.117 & $+$0.025 & $+$0.012 & $+$0.005 & $+$0.005 & $+$0.004 & $+$0.739 & $+$0.024 & $+$0.006 & $-$0.001 & $-$0.003 & $-$0.003 \\
 $-$0.947 & $-$0.038 & $-$0.015 & $-$0.005 & $-$0.001 & $-$0.008 & $-$0.091 & $+$0.018 & $+$0.012 & $+$0.005 & $+$0.003 & $+$0.002 & $+$0.765 & $+$0.015 & $+$0.002 & $-$0.003 & $-$0.003 & $-$0.005 \\
 $-$0.921 & $-$0.024 & $-$0.011 & $-$0.005 & $-$0.004 & $-$0.009 & $-$0.065 & $+$0.025 & $+$0.013 & $+$0.005 & $+$0.004 & $+$0.008 & $+$0.791 & $+$0.015 & $+$0.001 & $-$0.000 & $-$0.001 & $+$0.001 \\
 $-$0.895 & $-$0.035 & $-$0.012 & $-$0.003 & $+$0.000 & $-$0.002 & $-$0.039 & $+$0.025 & $+$0.012 & $+$0.007 & $+$0.004 & $+$0.011 & $+$0.817 & $+$0.014 & $+$0.005 & $+$0.001 & $-$0.000 & $+$0.002 \\
 $-$0.869 & $-$0.039 & $-$0.013 & $-$0.002 & $+$0.002 & $+$0.003 & $-$0.013 & $+$0.012 & $+$0.008 & $+$0.007 & $+$0.007 & $+$0.017 & $+$0.843 & $+$0.029 & $+$0.012 & $+$0.004 & $+$0.001 & $+$0.003 \\
 $-$0.843 & $-$0.034 & $-$0.014 & $-$0.004 & $-$0.001 & $+$0.004 & $+$0.013 & $-$0.006 & $-$0.003 & $-$0.003 & $-$0.004 & $+$0.001 & $+$0.869 & $+$0.016 & $+$0.007 & $+$0.007 & $+$0.009 & $+$0.010 \\
 $-$0.817 & $-$0.043 & $-$0.014 & $+$0.001 & $+$0.007 & $+$0.009 & $+$0.039 & $-$0.006 & $-$0.007 & $-$0.009 & $-$0.004 & $-$0.002 & $+$0.895 & $+$0.015 & $+$0.009 & $+$0.005 & $+$0.007 & $+$0.006 \\
 $-$0.791 & $-$0.026 & $-$0.005 & $+$0.003 & $+$0.004 & $+$0.005 & $+$0.065 & $-$0.002 & $-$0.006 & $-$0.010 & $-$0.009 & $-$0.007 & $+$0.921 & $+$0.000 & $+$0.007 & $+$0.007 & $+$0.008 & $+$0.005 \\
 $-$0.765 & $-$0.022 & $-$0.002 & $+$0.009 & $+$0.011 & $+$0.009 & $+$0.091 & $-$0.018 & $-$0.012 & $-$0.011 & $-$0.009 & $-$0.006 & $+$0.947 & $-$0.006 & $+$0.006 & $+$0.009 & $+$0.012 & $+$0.010 \\
 $-$0.739 & $-$0.007 & $+$0.002 & $+$0.005 & $+$0.008 & $+$0.005 & $+$0.117 & $-$0.017 & $-$0.010 & $-$0.010 & $-$0.012 & $-$0.013 & $+$0.973 & $-$0.001 & $+$0.008 & $+$0.010 & $+$0.014 & $+$0.013 \\
 $-$0.713 & $-$0.006 & $+$0.001 & $+$0.002 & $+$0.002 & $-$0.001 & $+$0.143 & $-$0.005 & $-$0.004 & $-$0.008 & $-$0.009 & $-$0.008 & $+$0.999 & $-$0.001 & $+$0.005 & $+$0.010 & $+$0.011 & $+$0.009 \\
 $-$0.687 & $-$0.006 & $-$0.003 & $+$0.003 & $+$0.005 & $+$0.005 & $+$0.169 & $-$0.007 & $-$0.003 & $-$0.004 & $-$0.007 & $+$0.000 & $+$1.025 & $-$0.012 & $-$0.001 & $+$0.004 & $+$0.005 & $+$0.005 \\
 $-$0.661 & $-$0.006 & $-$0.003 & $-$0.003 & $-$0.001 & $-$0.000 & $+$0.195 & $-$0.014 & $-$0.006 & $-$0.003 & $-$0.002 & $+$0.005 & $+$1.050 & $+$0.018 & $+$0.010 & $+$0.008 & $+$0.006 & $+$0.007 \\
 $-$0.635 & $-$0.016 & $-$0.013 & $-$0.011 & $-$0.009 & $-$0.008 & $+$0.220 & $-$0.020 & $-$0.009 & $-$0.002 & $-$0.001 & $+$0.005 & $+$1.076 & $+$0.019 & $+$0.007 & $+$0.002 & $+$0.004 & $+$0.003 \\
 $-$0.610 & $-$0.031 & $-$0.023 & $-$0.014 & $-$0.013 & $-$0.014 & $+$0.246 & $+$0.002 & $+$0.001 & $-$0.000 & $+$0.003 & $+$0.010 & $+$1.102 & $-$0.009 & $-$0.000 & $+$0.005 & $+$0.007 & $+$0.007 \\
 $-$0.584 & $-$0.032 & $-$0.017 & $-$0.013 & $-$0.010 & $-$0.008 & $+$0.272 & $+$0.006 & $+$0.003 & $-$0.001 & $-$0.001 & $+$0.001 & $+$1.128 & $+$0.002 & $+$0.006 & $+$0.006 & $+$0.005 & $+$0.001 \\
 $-$0.558 & $-$0.025 & $-$0.016 & $-$0.010 & $-$0.008 & $-$0.007 & $+$0.298 & $-$0.001 & $-$0.001 & $-$0.001 & $-$0.001 & $+$0.002 & $+$1.154 & $+$0.001 & $+$0.003 & $+$0.003 & $+$0.004 & $+$0.000 \\
 $-$0.532 & $-$0.020 & $-$0.015 & $-$0.011 & $-$0.007 & $-$0.009 & $+$0.324 & $+$0.013 & $+$0.003 & $-$0.000 & $-$0.003 & $-$0.004 & $+$1.180 & $+$0.006 & $+$0.011 & $+$0.011 & $+$0.011 & $+$0.008 \\
 $-$0.506 & $-$0.012 & $-$0.010 & $-$0.009 & $-$0.009 & $-$0.008 & $+$0.350 & $+$0.001 & $-$0.000 & $-$0.001 & $-$0.001 & $-$0.001 & $+$1.206 & $-$0.001 & $+$0.009 & $+$0.012 & $+$0.013 & $+$0.011 \\
 $-$0.480 & $-$0.025 & $-$0.014 & $-$0.008 & $-$0.005 & $-$0.004 & $+$0.376 & $+$0.001 & $+$0.001 & $-$0.002 & $-$0.003 & $-$0.001 & $+$1.232 & $-$0.018 & $+$0.002 & $+$0.010 & $+$0.013 & $+$0.011 \\
 $-$0.454 & $-$0.020 & $-$0.011 & $-$0.006 & $-$0.006 & $-$0.002 & $+$0.402 & $+$0.009 & $+$0.002 & $-$0.002 & $-$0.000 & $+$0.003 & $+$1.262 & $-$0.007 & $+$0.006 & $+$0.017 & $+$0.014 & $+$0.015 \\
 $-$0.428 & $+$0.012 & $+$0.000 & $-$0.005 & $-$0.007 & $-$0.004 & $+$0.428 & $+$0.035 & $+$0.012 & $+$0.004 & $+$0.000 & $+$0.001 \\

\hline
\end{tabular}
\end{table*}

\begin{table*}
\centering
\caption{$\delta_m^{ext}({\rm RA})$ of the north strip for the V4.2 catalog.}
\label{} \tiny 
\begin{tabular}{rrrrrrrrrrrrrrrrrr} \hline\hline
 RA   & $u$ &  $g$ & $r$  & $i$ & $z$ & RA & $u$ &  $g$ & $r$  & $i$ & $z$  & RA   & $u$ &  $g$ & $r$  & $i$ & $z$   \\
 (deg) & (mag) &  (mag) & (mag)  & (mag) & (mag) &(deg)   & (mag) &  (mag) & (mag)  & (mag) & (mag) & (deg)   & (mag) & (mag)& (mag) & (mag) & (mag)   \\\hline
 $-$50.5 & $-$0.017 & $-$0.002 & $+$0.000 & $-$0.004 & $-$0.001 & $-$13.5 & $-$0.008 & $+$0.001 & $+$0.001 & $+$0.002 & $+$0.001 & $+$23.5 & $-$0.001 & $+$0.002 & $-$0.001 & $+$0.003 & $+$0.005 \\
 $-$49.5 & $-$0.003 & $-$0.000 & $-$0.002 & $-$0.003 & $-$0.008 & $-$12.5 & $-$0.011 & $-$0.001 & $+$0.000 & $-$0.000 & $-$0.001 & $+$24.5 & $-$0.011 & $+$0.002 & $+$0.001 & $+$0.003 & $+$0.006 \\
 $-$48.5 & $-$0.002 & $+$0.005 & $+$0.002 & $-$0.004 & $-$0.008 & $-$11.5 & $-$0.012 & $-$0.003 & $-$0.000 & $-$0.001 & $+$0.002 & $+$25.5 & $+$0.003 & $+$0.006 & $+$0.001 & $+$0.002 & $+$0.007 \\
 $-$47.5 & $-$0.012 & $-$0.000 & $-$0.000 & $-$0.002 & $-$0.003 & $-$10.5 & $-$0.005 & $+$0.000 & $+$0.000 & $+$0.000 & $+$0.000 & $+$26.5 & $-$0.007 & $+$0.000 & $+$0.001 & $+$0.001 & $+$0.004 \\
 $-$46.5 & $-$0.021 & $-$0.001 & $+$0.001 & $-$0.001 & $-$0.002 & $-$9.5 & $+$0.002 & $+$0.001 & $+$0.002 & $-$0.001 & $-$0.001 & $+$27.5 & $-$0.005 & $+$0.000 & $+$0.001 & $+$0.002 & $+$0.005 \\
 $-$45.5 & $-$0.015 & $-$0.000 & $-$0.001 & $-$0.003 & $-$0.006 & $-$8.5 & $+$0.002 & $+$0.001 & $-$0.000 & $+$0.000 & $+$0.001 & $+$28.5 & $-$0.008 & $-$0.001 & $-$0.004 & $-$0.000 & $+$0.003 \\
 $-$44.5 & $-$0.006 & $+$0.004 & $+$0.001 & $-$0.003 & $-$0.005 & $-$7.5 & $-$0.002 & $-$0.000 & $-$0.001 & $-$0.001 & $+$0.002 & $+$29.5 & $+$0.004 & $+$0.002 & $-$0.000 & $+$0.002 & $+$0.006 \\
 $-$43.5 & $-$0.001 & $+$0.005 & $+$0.003 & $+$0.000 & $-$0.005 & $-$6.5 & $+$0.004 & $-$0.001 & $+$0.001 & $+$0.000 & $+$0.003 & $+$30.5 & $+$0.007 & $-$0.001 & $-$0.000 & $+$0.003 & $+$0.004 \\
 $-$42.5 & $-$0.012 & $+$0.001 & $+$0.004 & $+$0.000 & $-$0.001 & $-$5.5 & $+$0.002 & $-$0.001 & $-$0.003 & $+$0.000 & $+$0.001 & $+$31.5 & $-$0.006 & $+$0.001 & $-$0.002 & $+$0.002 & $+$0.003 \\
 $-$41.5 & $-$0.029 & $-$0.004 & $+$0.002 & $+$0.001 & $+$0.001 & $-$4.5 & $+$0.007 & $+$0.001 & $+$0.001 & $+$0.000 & $+$0.002 & $+$32.5 & $+$0.006 & $+$0.004 & $+$0.002 & $+$0.005 & $+$0.006 \\
 $-$40.5 & $-$0.012 & $+$0.002 & $-$0.000 & $-$0.002 & $-$0.004 & $-$3.5 & $+$0.002 & $+$0.002 & $+$0.000 & $+$0.001 & $+$0.003 & $+$33.5 & $+$0.003 & $+$0.002 & $-$0.000 & $+$0.002 & $+$0.005 \\
 $-$39.5 & $+$0.004 & $+$0.003 & $-$0.001 & $-$0.001 & $-$0.002 & $-$2.5 & $+$0.005 & $+$0.002 & $+$0.002 & $+$0.001 & $+$0.002 & $+$34.5 & $-$0.000 & $+$0.003 & $+$0.001 & $+$0.003 & $+$0.005 \\
 $-$38.5 & $-$0.008 & $+$0.003 & $-$0.001 & $-$0.001 & $+$0.003 & $-$1.5 & $-$0.004 & $-$0.001 & $-$0.003 & $-$0.002 & $+$0.003 & $+$35.5 & $-$0.001 & $+$0.002 & $-$0.000 & $+$0.002 & $+$0.005 \\
 $-$37.5 & $-$0.006 & $+$0.002 & $-$0.001 & $+$0.000 & $+$0.000 & $-$0.5 & $-$0.010 & $+$0.002 & $+$0.000 & $+$0.000 & $+$0.003 & $+$36.5 & $+$0.007 & $+$0.002 & $+$0.001 & $+$0.002 & $+$0.007 \\
 $-$36.5 & $-$0.003 & $+$0.003 & $-$0.001 & $+$0.000 & $-$0.000 & $+$0.5 & $-$0.008 & $-$0.001 & $+$0.000 & $+$0.002 & $+$0.004 & $+$37.5 & $+$0.003 & $+$0.002 & $+$0.001 & $+$0.003 & $+$0.008 \\
 $-$35.5 & $+$0.001 & $+$0.003 & $-$0.000 & $-$0.000 & $-$0.000 & $+$1.5 & $-$0.008 & $+$0.001 & $-$0.000 & $+$0.001 & $+$0.004 & $+$38.5 & $+$0.013 & $+$0.005 & $+$0.001 & $+$0.002 & $+$0.005 \\
 $-$34.5 & $-$0.007 & $+$0.000 & $-$0.000 & $-$0.003 & $-$0.005 & $+$2.5 & $-$0.015 & $-$0.004 & $-$0.001 & $-$0.000 & $+$0.002 & $+$39.5 & $+$0.001 & $+$0.002 & $-$0.001 & $+$0.002 & $+$0.005 \\
 $-$33.5 & $-$0.002 & $+$0.001 & $+$0.000 & $-$0.004 & $-$0.006 & $+$3.5 & $-$0.009 & $-$0.002 & $-$0.001 & $-$0.001 & $+$0.002 & $+$40.5 & $+$0.003 & $+$0.001 & $+$0.001 & $+$0.000 & $+$0.002 \\
 $-$32.5 & $-$0.012 & $+$0.001 & $+$0.003 & $+$0.000 & $-$0.000 & $+$4.5 & $-$0.019 & $-$0.002 & $-$0.000 & $-$0.001 & $+$0.002 & $+$41.5 & $+$0.001 & $-$0.002 & $-$0.000 & $+$0.000 & $+$0.002 \\
 $-$31.5 & $-$0.003 & $-$0.003 & $+$0.002 & $+$0.001 & $-$0.002 & $+$5.5 & $+$0.006 & $+$0.004 & $-$0.001 & $+$0.003 & $+$0.008 & $+$42.5 & $+$0.003 & $-$0.002 & $-$0.003 & $-$0.002 & $+$0.001 \\
 $-$30.5 & $-$0.011 & $-$0.002 & $+$0.000 & $-$0.002 & $-$0.003 & $+$6.5 & $+$0.010 & $+$0.004 & $-$0.001 & $-$0.000 & $+$0.000 & $+$43.5 & $+$0.002 & $-$0.002 & $-$0.002 & $-$0.001 & $-$0.000 \\
 $-$29.5 & $-$0.003 & $-$0.000 & $+$0.003 & $-$0.002 & $-$0.007 & $+$7.5 & $-$0.011 & $-$0.002 & $+$0.000 & $+$0.003 & $+$0.003 & $+$44.5 & $+$0.004 & $-$0.002 & $-$0.001 & $-$0.002 & $-$0.003 \\
 $-$28.5 & $-$0.002 & $+$0.004 & $+$0.002 & $+$0.002 & $-$0.001 & $+$8.5 & $-$0.004 & $+$0.001 & $+$0.000 & $+$0.001 & $+$0.002 & $+$45.5 & $+$0.001 & $-$0.002 & $-$0.001 & $-$0.002 & $+$0.000 \\
 $-$27.5 & $-$0.004 & $+$0.000 & $+$0.001 & $-$0.001 & $-$0.005 & $+$9.5 & $-$0.000 & $+$0.002 & $-$0.001 & $-$0.000 & $+$0.002 & $+$46.5 & $+$0.003 & $-$0.002 & $-$0.001 & $-$0.002 & $-$0.004 \\
 $-$26.5 & $-$0.008 & $+$0.003 & $+$0.000 & $+$0.001 & $-$0.001 & $+$10.5 & $-$0.005 & $-$0.001 & $+$0.000 & $+$0.002 & $+$0.003 & $+$47.5 & $+$0.001 & $-$0.002 & $-$0.002 & $-$0.001 & $-$0.002 \\
 $-$25.5 & $-$0.007 & $-$0.001 & $+$0.001 & $-$0.001 & $-$0.004 & $+$11.5 & $+$0.002 & $+$0.001 & $+$0.002 & $+$0.001 & $+$0.003 & $+$48.5 & $+$0.010 & $+$0.000 & $-$0.001 & $-$0.001 & $-$0.001 \\
 $-$24.5 & $-$0.004 & $+$0.000 & $+$0.001 & $-$0.001 & $-$0.004 & $+$12.5 & $+$0.003 & $+$0.005 & $-$0.000 & $+$0.000 & $+$0.003 & $+$49.5 & $+$0.008 & $-$0.001 & $-$0.001 & $-$0.001 & $-$0.002 \\
 $-$23.5 & $-$0.002 & $+$0.001 & $+$0.001 & $-$0.001 & $-$0.003 & $+$13.5 & $+$0.004 & $+$0.001 & $-$0.001 & $+$0.001 & $+$0.003 & $+$50.5 & $+$0.008 & $-$0.004 & $-$0.001 & $-$0.003 & $-$0.004 \\
 $-$22.5 & $-$0.014 & $-$0.004 & $+$0.002 & $-$0.000 & $-$0.006 & $+$14.5 & $+$0.008 & $+$0.002 & $+$0.003 & $+$0.003 & $+$0.005 & $+$51.5 & $+$0.004 & $-$0.003 & $-$0.001 & $-$0.000 & $+$0.001 \\
 $-$21.5 & $-$0.005 & $+$0.001 & $-$0.001 & $-$0.001 & $-$0.003 & $+$15.5 & $+$0.003 & $+$0.002 & $+$0.001 & $+$0.002 & $+$0.003 & $+$52.5 & $+$0.009 & $-$0.005 & $-$0.001 & $-$0.001 & $-$0.002 \\
 $-$20.5 & $+$0.004 & $-$0.001 & $+$0.001 & $-$0.001 & $-$0.003 & $+$16.5 & $-$0.003 & $-$0.002 & $-$0.001 & $-$0.000 & $+$0.002 & $+$53.5 & $+$0.008 & $-$0.003 & $-$0.001 & $-$0.003 & $-$0.003 \\
 $-$19.5 & $-$0.009 & $-$0.002 & $-$0.000 & $-$0.002 & $-$0.002 & $+$17.5 & $-$0.005 & $+$0.000 & $+$0.002 & $+$0.001 & $+$0.004 & $+$54.5 & $-$0.000 & $-$0.005 & $-$0.001 & $-$0.004 & $-$0.006 \\
 $-$18.5 & $-$0.002 & $-$0.002 & $-$0.001 & $-$0.001 & $-$0.003 & $+$18.5 & $-$0.004 & $-$0.002 & $-$0.001 & $+$0.001 & $+$0.003 & $+$55.5 & $-$0.003 & $-$0.008 & $-$0.003 & $-$0.005 & $-$0.006 \\
 $-$17.5 & $-$0.011 & $-$0.001 & $+$0.002 & $-$0.001 & $-$0.002 & $+$19.5 & $+$0.003 & $+$0.002 & $+$0.001 & $+$0.000 & $+$0.002 & $+$56.5 & $+$0.003 & $-$0.004 & $+$0.003 & $-$0.001 & $-$0.005 \\
 $-$16.5 & $-$0.011 & $-$0.008 & $-$0.002 & $-$0.002 & $+$0.000 & $+$20.5 & $+$0.006 & $+$0.003 & $+$0.001 & $+$0.002 & $+$0.005 & $+$57.5 & $+$0.016 & $-$0.004 & $+$0.007 & $+$0.004 & $+$0.005 \\
 $-$15.5 & $+$0.006 & $+$0.004 & $-$0.002 & $-$0.003 & $+$0.000 & $+$21.5 & $-$0.000 & $+$0.001 & $-$0.001 & $+$0.001 & $+$0.006 & $+$58.5 & $+$0.017 & $-$0.001 & $+$0.009 & $+$0.012 & $+$0.014 \\
 $-$14.5 & $-$0.002 & $-$0.003 & $-$0.003 & $-$0.002 & $+$0.000 & $+$22.5 & $-$0.003 & $-$0.001 & $-$0.002 & $+$0.001 & $+$0.003 & $+$59.5 & $+$0.014 & $+$0.007 & $+$0.009 & $+$0.017 & $+$0.021 \\

\hline
\end{tabular}
\end{table*}

\begin{table*}
\centering
\caption{$\delta_m^{ext}({\rm RA})$ of the south strip for the V4.2 catalog.}
\label{} \tiny 
\begin{tabular}{rrrrrrrrrrrrrrrrrr} \hline\hline
 RA   & $u$ &  $g$ & $r$  & $i$ & $z$ & RA & $u$ &  $g$ & $r$  & $i$ & $z$  & RA   & $u$ &  $g$ & $r$  & $i$ & $z$   \\
 (deg) & (mag) &  (mag) & (mag)  & (mag) & (mag) &(deg)   & (mag) &  (mag) & (mag)  & (mag) & (mag) & (deg)   & (mag) & (mag)& (mag) & (mag) & (mag)   \\\hline
 $-$50.5 & $+$0.004 & $+$0.002 & $+$0.001 & $-$0.002 & $-$0.003 & $-$13.5 & $-$0.011 & $-$0.000 & $-$0.001 & $+$0.001 & $-$0.003 & $+$23.5 & $-$0.008 & $+$0.002 & $+$0.001 & $+$0.002 & $+$0.002 \\
 $-$49.5 & $-$0.016 & $-$0.004 & $+$0.000 & $-$0.005 & $-$0.004 & $-$12.5 & $-$0.013 & $+$0.000 & $-$0.000 & $-$0.000 & $-$0.004 & $+$24.5 & $-$0.015 & $-$0.003 & $-$0.000 & $+$0.000 & $-$0.000 \\
 $-$48.5 & $-$0.017 & $+$0.003 & $+$0.002 & $-$0.003 & $-$0.003 & $-$11.5 & $-$0.003 & $+$0.002 & $+$0.003 & $+$0.002 & $+$0.001 & $+$25.5 & $-$0.001 & $+$0.005 & $+$0.002 & $+$0.002 & $+$0.002 \\
 $-$47.5 & $-$0.011 & $-$0.001 & $-$0.001 & $-$0.002 & $+$0.001 & $-$10.5 & $-$0.000 & $+$0.001 & $+$0.002 & $+$0.001 & $+$0.001 & $+$26.5 & $-$0.004 & $+$0.001 & $-$0.001 & $+$0.002 & $+$0.002 \\
 $-$46.5 & $-$0.009 & $-$0.002 & $-$0.001 & $-$0.004 & $-$0.004 & $-$9.5 & $+$0.009 & $+$0.004 & $+$0.004 & $+$0.001 & $-$0.000 & $+$27.5 & $+$0.003 & $+$0.001 & $-$0.001 & $+$0.001 & $-$0.001 \\
 $-$45.5 & $-$0.010 & $+$0.003 & $+$0.003 & $-$0.004 & $-$0.004 & $-$8.5 & $-$0.002 & $+$0.000 & $-$0.002 & $-$0.002 & $-$0.000 & $+$28.5 & $+$0.001 & $+$0.001 & $+$0.002 & $+$0.003 & $+$0.003 \\
 $-$44.5 & $-$0.006 & $+$0.002 & $-$0.000 & $-$0.003 & $-$0.003 & $-$7.5 & $+$0.003 & $+$0.004 & $+$0.002 & $+$0.000 & $+$0.000 & $+$29.5 & $+$0.004 & $+$0.002 & $+$0.001 & $+$0.004 & $+$0.004 \\
 $-$43.5 & $-$0.001 & $+$0.005 & $-$0.000 & $-$0.001 & $-$0.002 & $-$6.5 & $+$0.001 & $+$0.003 & $+$0.002 & $-$0.000 & $+$0.001 & $+$30.5 & $-$0.002 & $+$0.001 & $+$0.001 & $+$0.005 & $+$0.006 \\
 $-$42.5 & $-$0.017 & $+$0.001 & $+$0.003 & $-$0.001 & $-$0.001 & $-$5.5 & $+$0.010 & $+$0.001 & $-$0.001 & $+$0.001 & $+$0.002 & $+$31.5 & $+$0.010 & $+$0.006 & $+$0.001 & $+$0.008 & $+$0.007 \\
 $-$41.5 & $-$0.021 & $-$0.000 & $+$0.000 & $-$0.001 & $-$0.001 & $-$4.5 & $+$0.003 & $-$0.002 & $-$0.000 & $-$0.001 & $+$0.001 & $+$32.5 & $+$0.001 & $+$0.004 & $+$0.005 & $+$0.008 & $+$0.007 \\
 $-$40.5 & $-$0.023 & $+$0.000 & $-$0.000 & $-$0.003 & $-$0.007 & $-$3.5 & $+$0.012 & $+$0.004 & $+$0.000 & $+$0.002 & $+$0.001 & $+$33.5 & $+$0.003 & $+$0.002 & $+$0.000 & $+$0.003 & $+$0.005 \\
 $-$39.5 & $-$0.003 & $+$0.005 & $+$0.000 & $+$0.000 & $-$0.004 & $-$2.5 & $-$0.002 & $+$0.002 & $+$0.002 & $+$0.000 & $+$0.004 & $+$34.5 & $+$0.006 & $+$0.004 & $+$0.002 & $+$0.005 & $+$0.007 \\
 $-$38.5 & $-$0.008 & $+$0.003 & $-$0.001 & $-$0.001 & $+$0.004 & $-$1.5 & $-$0.005 & $-$0.003 & $-$0.002 & $+$0.000 & $+$0.001 & $+$35.5 & $+$0.001 & $+$0.002 & $-$0.000 & $+$0.005 & $+$0.006 \\
 $-$37.5 & $-$0.010 & $+$0.001 & $-$0.001 & $-$0.001 & $-$0.000 & $-$0.5 & $+$0.001 & $+$0.003 & $-$0.000 & $-$0.002 & $+$0.000 & $+$36.5 & $-$0.002 & $+$0.001 & $+$0.000 & $+$0.003 & $+$0.005 \\
 $-$36.5 & $-$0.011 & $+$0.001 & $-$0.001 & $-$0.002 & $+$0.000 & $+$0.5 & $+$0.003 & $+$0.002 & $+$0.002 & $+$0.000 & $+$0.003 & $+$37.5 & $+$0.005 & $+$0.003 & $+$0.001 & $+$0.004 & $+$0.005 \\
 $-$35.5 & $-$0.011 & $+$0.002 & $-$0.002 & $-$0.002 & $+$0.001 & $+$1.5 & $-$0.006 & $+$0.001 & $+$0.002 & $+$0.000 & $-$0.001 & $+$38.5 & $+$0.003 & $+$0.004 & $+$0.000 & $+$0.003 & $+$0.002 \\
 $-$34.5 & $-$0.008 & $+$0.001 & $-$0.000 & $-$0.003 & $-$0.004 & $+$2.5 & $-$0.001 & $+$0.003 & $+$0.002 & $+$0.000 & $-$0.000 & $+$39.5 & $-$0.005 & $+$0.000 & $+$0.001 & $+$0.004 & $+$0.005 \\
 $-$33.5 & $+$0.001 & $-$0.000 & $-$0.001 & $-$0.004 & $-$0.007 & $+$3.5 & $-$0.008 & $-$0.002 & $-$0.001 & $-$0.001 & $+$0.003 & $+$40.5 & $+$0.002 & $+$0.000 & $-$0.002 & $+$0.001 & $+$0.001 \\
 $-$32.5 & $-$0.005 & $+$0.001 & $+$0.002 & $+$0.002 & $+$0.000 & $+$4.5 & $-$0.008 & $+$0.002 & $+$0.002 & $+$0.002 & $+$0.002 & $+$41.5 & $-$0.002 & $-$0.002 & $-$0.001 & $+$0.001 & $+$0.002 \\
 $-$31.5 & $-$0.010 & $-$0.001 & $+$0.002 & $+$0.001 & $+$0.002 & $+$5.5 & $+$0.001 & $+$0.003 & $-$0.000 & $+$0.001 & $+$0.005 & $+$42.5 & $+$0.005 & $+$0.000 & $-$0.003 & $+$0.000 & $-$0.001 \\
 $-$30.5 & $-$0.002 & $+$0.002 & $-$0.001 & $-$0.003 & $-$0.003 & $+$6.5 & $+$0.006 & $+$0.004 & $+$0.001 & $+$0.002 & $+$0.005 & $+$43.5 & $-$0.003 & $-$0.004 & $-$0.003 & $-$0.003 & $-$0.004 \\
 $-$29.5 & $-$0.014 & $-$0.000 & $+$0.002 & $-$0.002 & $-$0.005 & $+$7.5 & $+$0.001 & $+$0.001 & $-$0.000 & $+$0.002 & $+$0.003 & $+$44.5 & $+$0.004 & $-$0.001 & $-$0.002 & $-$0.002 & $-$0.006 \\
 $-$28.5 & $+$0.003 & $+$0.004 & $+$0.000 & $-$0.000 & $+$0.001 & $+$8.5 & $-$0.002 & $+$0.001 & $+$0.002 & $+$0.002 & $+$0.005 & $+$45.5 & $-$0.002 & $-$0.003 & $-$0.002 & $-$0.001 & $-$0.003 \\
 $-$27.5 & $-$0.002 & $+$0.002 & $+$0.002 & $+$0.000 & $-$0.002 & $+$9.5 & $+$0.001 & $-$0.001 & $+$0.000 & $+$0.001 & $+$0.005 & $+$46.5 & $-$0.002 & $-$0.004 & $-$0.004 & $-$0.003 & $-$0.006 \\
 $-$26.5 & $+$0.002 & $+$0.001 & $-$0.001 & $-$0.002 & $-$0.003 & $+$10.5 & $-$0.007 & $+$0.000 & $+$0.001 & $+$0.001 & $+$0.004 & $+$47.5 & $+$0.008 & $-$0.002 & $-$0.003 & $-$0.002 & $-$0.004 \\
 $-$25.5 & $-$0.004 & $+$0.001 & $+$0.001 & $-$0.001 & $-$0.003 & $+$11.5 & $-$0.003 & $+$0.000 & $+$0.002 & $+$0.002 & $+$0.005 & $+$48.5 & $+$0.008 & $-$0.003 & $-$0.002 & $-$0.002 & $-$0.004 \\
 $-$24.5 & $-$0.002 & $+$0.000 & $+$0.001 & $-$0.003 & $-$0.003 & $+$12.5 & $-$0.002 & $+$0.002 & $+$0.001 & $+$0.004 & $+$0.005 & $+$49.5 & $+$0.010 & $-$0.003 & $-$0.002 & $-$0.002 & $-$0.007 \\
 $-$23.5 & $+$0.004 & $+$0.001 & $+$0.001 & $-$0.002 & $-$0.002 & $+$13.5 & $+$0.008 & $+$0.001 & $+$0.003 & $+$0.003 & $+$0.005 & $+$50.5 & $+$0.005 & $-$0.006 & $-$0.004 & $-$0.005 & $-$0.008 \\
 $-$22.5 & $-$0.002 & $-$0.001 & $-$0.002 & $-$0.000 & $-$0.006 & $+$14.5 & $+$0.015 & $+$0.003 & $+$0.002 & $+$0.003 & $+$0.002 & $+$51.5 & $+$0.009 & $-$0.005 & $-$0.003 & $-$0.004 & $-$0.005 \\
 $-$21.5 & $-$0.003 & $+$0.001 & $+$0.002 & $-$0.002 & $-$0.005 & $+$15.5 & $+$0.004 & $+$0.002 & $+$0.001 & $+$0.003 & $+$0.004 & $+$52.5 & $+$0.002 & $-$0.006 & $-$0.003 & $-$0.004 & $-$0.006 \\
 $-$20.5 & $-$0.006 & $+$0.002 & $+$0.003 & $-$0.001 & $-$0.003 & $+$16.5 & $-$0.000 & $+$0.001 & $+$0.002 & $+$0.004 & $+$0.003 & $+$53.5 & $-$0.000 & $-$0.008 & $-$0.004 & $-$0.005 & $-$0.008 \\
 $-$19.5 & $+$0.000 & $+$0.001 & $+$0.001 & $-$0.001 & $-$0.000 & $+$17.5 & $-$0.006 & $-$0.001 & $+$0.002 & $+$0.003 & $+$0.002 & $+$54.5 & $-$0.003 & $-$0.008 & $-$0.004 & $-$0.003 & $-$0.007 \\
 $-$18.5 & $-$0.001 & $-$0.000 & $+$0.003 & $-$0.001 & $-$0.003 & $+$18.5 & $+$0.006 & $+$0.001 & $+$0.002 & $+$0.004 & $+$0.003 & $+$55.5 & $+$0.011 & $-$0.008 & $-$0.004 & $-$0.006 & $-$0.011 \\
 $-$17.5 & $-$0.013 & $-$0.003 & $+$0.001 & $-$0.000 & $-$0.004 & $+$19.5 & $-$0.001 & $-$0.000 & $+$0.002 & $+$0.004 & $+$0.002 & $+$56.5 & $+$0.006 & $-$0.009 & $-$0.000 & $-$0.002 & $-$0.004 \\
 $-$16.5 & $+$0.004 & $+$0.001 & $+$0.001 & $-$0.001 & $-$0.005 & $+$20.5 & $-$0.006 & $+$0.001 & $+$0.001 & $+$0.001 & $+$0.001 & $+$57.5 & $+$0.006 & $-$0.006 & $+$0.003 & $-$0.002 & $-$0.005 \\
 $-$15.5 & $+$0.021 & $+$0.008 & $+$0.004 & $-$0.002 & $-$0.006 & $+$21.5 & $-$0.003 & $+$0.003 & $+$0.002 & $+$0.003 & $+$0.003 & $+$58.5 & $+$0.015 & $-$0.004 & $+$0.002 & $+$0.009 & $+$0.013 \\
 $-$14.5 & $-$0.001 & $-$0.000 & $-$0.003 & $-$0.001 & $-$0.003 & $+$22.5 & $-$0.009 & $-$0.001 & $-$0.000 & $-$0.000 & $+$0.002 & $+$59.5 & $-$0.005 & $+$0.006 & $+$0.007 & $+$0.019 & $+$0.021 \\

\hline
\end{tabular}
\end{table*}

\begin{table*}
\centering
\caption{$\delta_m^{ff}({\rm Dec})$ for the V4.2 catalog.}
\label{} \tiny 
\begin{tabular}{rrrrrrrrrrrrrrrrrr} \hline\hline
 Dec   & $u$ &  $g$ & $r$  & $i$ & $z$ & Dec & $u$ &  $g$ & $r$  & $i$ & $z$  & Dec   & $u$ &  $g$ & $r$  & $i$ & $z$   \\
 (deg) & (mag) &  (mag) & (mag)  & (mag) & (mag) &(deg)   & (mag) &  (mag) & (mag)  & (mag) & (mag) & (deg)   & (mag) & (mag)& (mag) & (mag) & (mag)
 \\\hline
 $-$1.262 & $-$0.000 & $+$0.003 & $+$0.002 & $+$0.002 & $+$0.002 & $-$0.402 & $+$0.006 & $-$0.000 & $+$0.000 & $-$0.002 & $+$0.000 & $+$0.454 & $+$0.001 & $+$0.002 & $-$0.000 & $-$0.001 & $+$0.000 \\
 $-$1.232 & $-$0.012 & $-$0.005 & $-$0.001 & $+$0.001 & $+$0.002 & $-$0.376 & $+$0.000 & $-$0.000 & $+$0.000 & $-$0.001 & $+$0.001 & $+$0.480 & $+$0.005 & $+$0.002 & $+$0.000 & $-$0.000 & $+$0.002 \\
 $-$1.206 & $-$0.004 & $-$0.004 & $-$0.000 & $+$0.001 & $-$0.001 & $-$0.350 & $+$0.003 & $+$0.002 & $-$0.000 & $+$0.000 & $+$0.000 & $+$0.506 & $+$0.005 & $+$0.001 & $+$0.001 & $+$0.000 & $+$0.001 \\
 $-$1.180 & $-$0.005 & $-$0.002 & $+$0.000 & $-$0.000 & $-$0.003 & $-$0.324 & $+$0.001 & $+$0.001 & $+$0.001 & $+$0.000 & $+$0.000 & $+$0.532 & $-$0.002 & $+$0.001 & $+$0.001 & $+$0.001 & $+$0.002 \\
 $-$1.154 & $-$0.001 & $-$0.002 & $-$0.001 & $+$0.001 & $+$0.001 & $-$0.298 & $+$0.003 & $+$0.002 & $-$0.000 & $+$0.001 & $+$0.001 & $+$0.558 & $+$0.007 & $+$0.001 & $+$0.000 & $-$0.000 & $+$0.001 \\
 $-$1.128 & $-$0.005 & $-$0.005 & $-$0.000 & $+$0.002 & $-$0.000 & $-$0.272 & $+$0.005 & $+$0.002 & $+$0.001 & $+$0.000 & $+$0.000 & $+$0.584 & $-$0.001 & $+$0.001 & $+$0.000 & $+$0.000 & $+$0.002 \\
 $-$1.102 & $-$0.002 & $-$0.001 & $-$0.000 & $-$0.000 & $-$0.001 & $-$0.246 & $+$0.003 & $+$0.003 & $+$0.001 & $-$0.000 & $-$0.000 & $+$0.610 & $+$0.005 & $+$0.002 & $+$0.001 & $-$0.001 & $+$0.001 \\
 $-$1.076 & $-$0.005 & $-$0.001 & $-$0.001 & $+$0.001 & $+$0.000 & $-$0.220 & $-$0.005 & $+$0.003 & $+$0.002 & $+$0.000 & $+$0.000 & $+$0.635 & $-$0.000 & $+$0.002 & $+$0.001 & $+$0.001 & $-$0.001 \\
 $-$1.050 & $-$0.008 & $-$0.000 & $-$0.002 & $-$0.001 & $-$0.001 & $-$0.195 & $-$0.003 & $+$0.001 & $+$0.001 & $+$0.003 & $+$0.000 & $+$0.661 & $-$0.002 & $+$0.002 & $-$0.000 & $+$0.000 & $+$0.000 \\
 $-$1.025 & $-$0.001 & $+$0.001 & $+$0.000 & $+$0.002 & $-$0.001 & $-$0.169 & $+$0.004 & $+$0.001 & $+$0.001 & $+$0.002 & $+$0.000 & $+$0.687 & $-$0.003 & $+$0.002 & $-$0.001 & $-$0.000 & $-$0.001 \\
 $-$0.999 & $-$0.003 & $-$0.001 & $-$0.000 & $+$0.001 & $+$0.001 & $-$0.143 & $+$0.006 & $+$0.002 & $-$0.000 & $+$0.000 & $-$0.003 & $+$0.713 & $-$0.003 & $+$0.002 & $+$0.000 & $+$0.001 & $-$0.001 \\
 $-$0.973 & $+$0.002 & $-$0.000 & $-$0.001 & $+$0.001 & $-$0.000 & $-$0.117 & $+$0.009 & $+$0.003 & $+$0.001 & $+$0.001 & $-$0.002 & $+$0.739 & $-$0.012 & $+$0.001 & $-$0.000 & $+$0.001 & $+$0.000 \\
 $-$0.947 & $-$0.001 & $-$0.001 & $-$0.001 & $+$0.002 & $+$0.001 & $-$0.091 & $+$0.001 & $+$0.005 & $+$0.001 & $+$0.000 & $-$0.002 & $+$0.765 & $-$0.007 & $-$0.002 & $-$0.002 & $+$0.000 & $+$0.000 \\
 $-$0.921 & $+$0.003 & $-$0.001 & $-$0.001 & $+$0.001 & $+$0.000 & $-$0.065 & $+$0.004 & $+$0.003 & $+$0.001 & $+$0.000 & $-$0.002 & $+$0.791 & $-$0.001 & $-$0.001 & $-$0.001 & $-$0.000 & $-$0.000 \\
 $-$0.895 & $+$0.002 & $-$0.000 & $-$0.001 & $+$0.001 & $-$0.000 & $-$0.039 & $+$0.008 & $+$0.003 & $-$0.000 & $-$0.000 & $-$0.001 & $+$0.817 & $+$0.007 & $+$0.000 & $-$0.002 & $-$0.003 & $-$0.002 \\
 $-$0.869 & $-$0.002 & $-$0.001 & $-$0.000 & $+$0.001 & $+$0.002 & $-$0.013 & $+$0.001 & $+$0.001 & $+$0.001 & $-$0.001 & $-$0.001 & $+$0.843 & $+$0.009 & $+$0.002 & $+$0.000 & $+$0.000 & $+$0.004 \\
 $-$0.843 & $-$0.000 & $-$0.001 & $-$0.001 & $-$0.000 & $+$0.001 & $+$0.013 & $-$0.012 & $-$0.002 & $-$0.001 & $-$0.001 & $-$0.003 & $+$0.869 & $+$0.005 & $-$0.001 & $-$0.001 & $+$0.000 & $+$0.005 \\
 $-$0.817 & $-$0.001 & $+$0.000 & $+$0.002 & $+$0.003 & $-$0.000 & $+$0.039 & $-$0.010 & $-$0.003 & $-$0.003 & $-$0.003 & $-$0.005 & $+$0.895 & $+$0.008 & $-$0.000 & $-$0.001 & $-$0.001 & $+$0.001 \\
 $-$0.791 & $+$0.003 & $+$0.001 & $+$0.002 & $+$0.001 & $-$0.001 & $+$0.065 & $-$0.007 & $-$0.002 & $-$0.003 & $-$0.002 & $-$0.004 & $+$0.921 & $+$0.007 & $+$0.001 & $+$0.000 & $+$0.000 & $+$0.001 \\
 $-$0.765 & $+$0.000 & $+$0.001 & $+$0.003 & $+$0.002 & $-$0.001 & $+$0.091 & $-$0.007 & $-$0.004 & $-$0.003 & $-$0.002 & $-$0.003 & $+$0.947 & $+$0.007 & $+$0.001 & $+$0.000 & $+$0.001 & $+$0.002 \\
 $-$0.739 & $+$0.004 & $+$0.001 & $+$0.000 & $+$0.001 & $-$0.000 & $+$0.117 & $-$0.010 & $-$0.002 & $-$0.003 & $-$0.002 & $-$0.005 & $+$0.973 & $+$0.005 & $-$0.000 & $-$0.000 & $+$0.001 & $+$0.003 \\
 $-$0.713 & $+$0.002 & $+$0.001 & $+$0.002 & $+$0.002 & $-$0.002 & $+$0.143 & $-$0.009 & $-$0.001 & $-$0.002 & $-$0.001 & $-$0.005 & $+$0.999 & $+$0.009 & $+$0.002 & $+$0.001 & $+$0.001 & $+$0.003 \\
 $-$0.687 & $+$0.004 & $+$0.001 & $+$0.002 & $+$0.001 & $+$0.001 & $+$0.169 & $-$0.011 & $-$0.002 & $-$0.002 & $-$0.003 & $-$0.003 & $+$1.025 & $+$0.001 & $+$0.000 & $+$0.001 & $+$0.001 & $+$0.004 \\
 $-$0.661 & $+$0.004 & $+$0.001 & $+$0.002 & $+$0.001 & $+$0.000 & $+$0.195 & $-$0.013 & $-$0.003 & $-$0.002 & $-$0.003 & $-$0.003 & $+$1.050 & $+$0.002 & $+$0.001 & $+$0.001 & $+$0.001 & $+$0.005 \\
 $-$0.635 & $+$0.006 & $+$0.001 & $+$0.000 & $+$0.001 & $-$0.000 & $+$0.220 & $-$0.008 & $-$0.002 & $-$0.000 & $-$0.001 & $-$0.002 & $+$1.076 & $-$0.000 & $-$0.001 & $+$0.000 & $-$0.000 & $+$0.003 \\
 $-$0.610 & $+$0.003 & $-$0.003 & $+$0.000 & $-$0.000 & $-$0.003 & $+$0.246 & $-$0.004 & $-$0.002 & $-$0.002 & $-$0.002 & $+$0.000 & $+$1.102 & $-$0.003 & $-$0.001 & $+$0.000 & $-$0.000 & $+$0.004 \\
 $-$0.584 & $+$0.002 & $-$0.001 & $+$0.000 & $-$0.001 & $-$0.003 & $+$0.272 & $-$0.006 & $-$0.001 & $-$0.002 & $-$0.002 & $-$0.001 & $+$1.128 & $+$0.005 & $+$0.001 & $-$0.000 & $+$0.000 & $+$0.003 \\
 $-$0.558 & $+$0.003 & $-$0.003 & $-$0.000 & $-$0.001 & $-$0.003 & $+$0.298 & $-$0.002 & $-$0.001 & $-$0.001 & $-$0.001 & $-$0.001 & $+$1.154 & $+$0.003 & $-$0.000 & $-$0.000 & $+$0.000 & $+$0.004 \\
 $-$0.532 & $+$0.004 & $-$0.001 & $+$0.000 & $-$0.001 & $-$0.004 & $+$0.324 & $+$0.002 & $-$0.000 & $-$0.001 & $-$0.002 & $-$0.000 & $+$1.180 & $+$0.007 & $+$0.001 & $+$0.002 & $+$0.001 & $+$0.006 \\
 $-$0.506 & $+$0.005 & $-$0.001 & $-$0.000 & $-$0.001 & $-$0.004 & $+$0.350 & $-$0.005 & $-$0.001 & $-$0.001 & $-$0.002 & $-$0.002 & $+$1.206 & $+$0.003 & $+$0.001 & $+$0.002 & $+$0.001 & $+$0.006 \\
 $-$0.480 & $+$0.004 & $-$0.002 & $+$0.001 & $-$0.001 & $-$0.004 & $+$0.376 & $-$0.014 & $-$0.002 & $+$0.000 & $-$0.001 & $-$0.001 & $+$1.232 & $+$0.002 & $-$0.001 & $+$0.002 & $+$0.001 & $+$0.006 \\
 $-$0.454 & $+$0.003 & $-$0.001 & $+$0.001 & $+$0.000 & $-$0.003 & $+$0.402 & $-$0.010 & $+$0.000 & $-$0.001 & $-$0.001 & $+$0.000 & $+$1.262 & $-$0.001 & $-$0.000 & $+$0.004 & $-$0.000 & $+$0.009 \\
 $-$0.428 & $-$0.001 & $-$0.002 & $-$0.001 & $-$0.000 & $-$0.003 & $+$0.428 & $+$0.010 & $+$0.005 & $+$0.001 & $-$0.000 & $+$0.001  \\

\hline
\end{tabular}
\end{table*}

\end{document}